# *h*-ErMnO$_3$ absorbance, reflectivity and, emissivity in the THz to mid-infrared from 2 K to 1700 K: carrier screening, Fröhlich resonance, small polarons, and bipolarons

Néstor E. Massa,*,[1] Leire del Campo,[2] Karsten Holldack,[3] Aurélien Canizarès,[2] Vinh Ta Phuoc,[4] Paula Kayser,[5] and José Antonio Alonso[6]

[1] Centro CEQUINOR, Consejo Nacional de Investigaciones Científicas y Técnicas, Universidad Nacional de La Plata, Blvd. 120 1465, B1904 La Plata, Argentina.

[2] Centre National de la Recherche Scientifique, CEMHTI UPR3079, Université Orléans, F-45071 Orléans, France

[3] Helmholtz-Zentrum für Materialien und Energie GmbH,) Albert Einstein Str. 15, D-12489 Berlin, Germany.

[4] Groupement de Recherche Matériaux Microélectronique Acoustique Nanotechnologies-Université François Rabelais Tours, Faculté des Sciences & Techniques, F- 37200 Tours, France.

[5] Centre for Science at Extreme Conditions and School of Chemistry, University of Edinburgh, Kings Buildings, Mayfield Road, EH9 3FD Edinburgh, United Kingdom.

[6] Instituto de Ciencia de Materiales de Madrid, CSIC, Cantoblanco, E-28049 Madrid, Spain.

•e-mail: neemmassa@gmail.com

# ABSTRACT

We report the temperature dependent THz to mid-infrared response of hexagonal-ErMnO$_3$ using absorption, reflectivity, and emissivity techniques from 2 K to 1700 K. At low temperatures, lowest frequency vibrational modes which extend up the lock-in ferroelectric temperature, coexist with paramagnon excitations and are associated with well-defined crystal field Rare Earth pure magnetic replicas in an intriguing phonon-magnetic convergence enhancing the multiferroic character of *h*-ErMnO$_3$. Increasing the temperature, a number of vibrational bands close to the space group predicted undergo profile broadening and softening. In particular, a distinctive set of bands in the 288-329 cm$^{-1}$ (300 K) range has a component whose profile is carrier screened becoming nearly fully blurred in the intermediate phase between ~830 K and ~1500 K. Below T$_C$ ~830 K this asymmetric band, having one component still partially screened, further splits as spin phonon interaction and the tripling of the unit cell takes place revealing at T$_N$ ~79 K a delicate balance of long- and short-range interactions. Ambient Raman scattering brings up evidence of a Fröhlich resonance due to Coulomb interactions between carriers and the macroscopic field linked to the corresponding longitudinal optical phonon mode. We found it is dynamically correlated to the hexagonal *c*-axis negative thermal expansion. Quantitative analyses of the mid-infrared (MIR) optical conductivity show that also plays a role in small polarons and mediates in high temperature bipolarones.

Bipolaron profiles at high temperatures change as the sample opacity increases when at ~900 K straight stripes turn curly toward complex vortex-antivortex domain patterns in the paraelectric phase. At still higher temperatures a low frequency Drude contribution is triggered by electron hopping signaling an insulator-metal phase transition at ~1600 K while the MIR response suggests coexistence between single small polarons and bipolarons. On closing, we draw a parallel with improper ferroelectrics sustaining a lattice incommensurate intermediate phase and unit cell tripling. We argue that in the *h*-RMnO$_3$ (R=Rare Earth, Y) family of compounds the intermediate phase be considered incommensurate with onset at T$_{INC}$ ~1500 K and ferroelectric lock-in at T$_C$ ~830 K delimiting this regime in *h*-ErMnO$_3$.

1. Research Areas
   Lattice dynamics Phonons **Polarons** Optical Properties
1. Physical Systems
   Strongly correlated systems
1. Techniques
   Fourier transform infrared spectroscopy  Infrared spectroscopy   Raman scattering

# INTRODUCTION

Multiferroics are a family of materials in which degrees of magnetic and electronic ordering coexist in a common temperature range leading in an ordered phase to control of one property by the other by magnetoelectric correlations.[1,2] Their lattices, in which spin, charge, and phonon arrangements interplay supporting multiferroicity and/or magnetoelectricity, are regularly built by a transition metal and a Rare Earth in oxygen cages.[3,4] They comprise perovskite distorted lattices undergoing changes eventually leading to stable hexagonal phases that are marginally related to the ideal cubic perovskite. From them, $RMnO_3$, where R is a Rare Earth or Y, is a distinctive family of compounds sharing properties in topologically related orthorhombic and hexagonal phases. The orthorhombic perovskite arrangement (typically described by the space group P*bnm*) is no longer stable for Rare-Earth cations smaller than $Tb^{3+}$ since a hexagonal non-perovskite phase with the same stoichiometry strongly competes in stability,[5] being the ferroelectric structure determined by electrostatics and covalency interactions.[6,7]

Hexagonal manganites have their 3+ transition metal located at the center of a $MnO_5$ bipyramids, trigonal bipyramid coordination identified as hexahedra bi-prims, having 5-fold oxygen coordination with no $Mn^{3+}$ Jahn-Teller distortion.[8] Among these manganites, $ErMnO_3$ is at the borderline between an orthorhombic O-$ErMnO_3$ distorted perovskite and the lattice structure where decrease of the ionic radius, at the A site, turns the orthorhombic phase metastable making more competitive a non-perovskite *h*-$ErMnO_3$ hexagonal phase. This yields a non-centrosymmetric lattice belonging to the space group $P6_3cm(C^3_{6v})$ $Z=6$)[5,9] allowing off-center displacements concomitant with ferroelectricity along the <u>*c*</u> axis,[4] and a relative high Curie temperature $T_C$ ~830 K.[10] Between this temperature and ~1500 K, above which becomes paraelectric, there is an intermediate phase with residual polarization in which on cooling the bipyramid weakly coupled planes gradually become corrugated by bipyramid tilting leading to low temperature unit cell tripling.

The planes house intraplane frustrated magnetic six-fold axis triangular arrangements with Mn moments at ∼120° (**Refs. 11, 12**) The antiferromagnetic order sets below $T_N$ at ~79 K in the $\Gamma_4$ representation[13,14] They are coupled antiferromagnetically along the *c* by Mn—O—O—Mn interplay exchange.[15]

Although studies of the lower temperature ferroelectric phase have been reported for *h*-ErMnO$_3$ phonons[16,17] and THz magnetic optical activity,[18,19] there is no study offering a complete view in the full range of existence yielding a global and better understanding of the electrodynamics of this compound. Here, we report the evolution from low temperatures increasing up to sample decomposition, of main magnetoelectric, lattice, small polaron and bipolaron features from 5 cm$^{-1}$ to 13000 cm$^{-1}$, i.e., in the sub-THz to mid/near-infrared (MIR to NIR) spectral region using absorption, reflection, and emission spectroscopic methods.

We found in *h*-ErMnO$_3$ strong magnetic field dependences assigned to M$^{3+}$ moments from 2 K to 15 K, in addition to the Er$^{3+}$ manifold peaking,[18] earlier identified as dielectric anomalies due magnetoelectric couplings,[20] in the same spectral region of a far infrared reflectivity tail amounting coexistence of hybrid excitations.

From 12 K to 300 K, phonon profiles follow the expected for nominal softening and thermal broadening as evidence of strong coupling and anharmonicities. A vibrational mode, with reststrahlen at ~300 cm$^{-1}$, undergoes anomalous screening by delocalized carriers that may be assimilated to due to Mott electron correlations playing on cooling a polar role, beyond the ion displacements of the non-centrosymmetric *P6$_3$cm(C$^3_{6v}$)* ferroelectric phase, as the system goes from a Mott-like insulator to an ionic-like insulator in a de-facto electronic phase transition at T$_C$.[21] Mode screening suggests strong electron-phonon interactions known for electrons coupling to longitudinal polar modes. We verified this hypothesis using Raman techniques exciting with above the gap the λ$_{exc}$= 355 nm line. At 300 K a red shifted Fröhlich resonance,[22] a sharp band peaking at 322 cm$^{-1}$, appears at mid-frequency on the infrared macroscopic field plateau associated to the LO ~330 cm$^{-1}$ mode.

Our temperature dependent mid-infrared (MIR) transmission, reflectivity, and emission also corroborate that strong carrier-LO interactions modes yield in *h*-ErMnO$_3$ the fingerprint of quasiparticles named polarons,[23,24], and, more specifically, bipolarons.[25]

Above T$_{INC}$~1500 K, the sample gains in opacity as stripes turn into the more complex patterns of the paraelectric centrosymmetric P6$_3$/mmc–D$_{6h}^4$(Z=2) hexagonal lattice.[26] Increasing the temperature even further, we find the onset of a weak Drude tail marking a metal-insulator transition at T$_{MI}$~1600 K due to hopping electrons in what may be interpreted as early stages of compound decomposition.

We conclude drawing similarities between $h$-ErMnO$_3$ and the family of improper ferroelectrics, K$_2$SeO$_4$ and X$_2$ZnCl$_4$ (K, Rb), that have a lattice incommensurate intermediate phase,[27] and suggesting that the delimiting temperatures naming phase transitions at ~1500 K and ~830 K should be identified T$_{INC}$ and T$_C$, respectively, determining the onset of the lattice incommensurate and the "locking-in" in the lower temperature polar phase.

## Experimental details

Our $h$-ErMnO$_3$ samples, prepared as polycrystalline powder by a liquid-mix technique, were characterized by X-rays and high-resolution neutron diffraction patterns (NDP) which assess their quality as well crystallized samples with perfectly modeling crystal structures.[28] NDP data were collected at ILL-Grenoble, in the high-flux D20 diffractometer with a wavelength λ= 1.313 Å. As shown in Fig. 1(a), at ambient temperature each Mn atom is coordinated by five oxygen atoms in a bipyramidal configuration. One O3 atom and two O4 atoms are in the equatorial plane of the bipyramid, whereas the O1 and O2 atoms are at the apexes. Er occupies two crystallographic positions, each bonded to seven oxygen atoms along the **c** axis. The structure for $h$-ErMnO$_3$ consists of layers of corner-sharing MnO$_5$ bipyramids separated by layers of edge-sharing ErO$_7$ polyhedra.[29] Fig. 1c displays the quality of the fit from NPD data, and the inset contains a view along the $c$ axis of one layer of MnO$_5$ units sharing corners forming a characteristic triangular arrangement.

Low temperature-low frequency absorbance measurements from 3 cm$^{-1}$ to 50 cm$^{-1}$ with 0.5 cm$^{-1}$ resolution have been performed at the THz beamline of the BESSY II storage ring at the Helmholtz-Zentrum Berlin (HZB) in the low-alpha multi bunch hybrid mode. Measurements in the 30 cm$^{-1}$ to 300 cm$^{-1}$ range were also taken at the beamline using the internal source of a Bruker IFS125 HR spectrometer. In every case a superconducting magnet (Oxford Spectromag 4000, 10T (here up to 7T) interfaced with the interferometer was used for the measurements under magnetic fields.[30]

Our preliminary Raman data were taken at 1 cm$^{-1}$ resolution with a Qontor l spectrometer equipped with an Leica 100x microscope objective (NA 0.85) with a 1800 g/mm grating with the 457 nm (Coherent Sapphire single frequency) and 545 nm (Cobolt Fandango™) exciting lines. The same set up was used with $\lambda_{exc}$ =355nm but replacing the 1800 gr/mm by the 3600 gr/mm grating and a Thorlabs LMU-40X-NUV (NA 0.5) focusing objective. The triple subtractive configuration (1800 g/mm gratings) of the Horiba Jobin Yvon T64000 spectrometer allowed Raman acquisition in the low-frequency range up to ~100 cm$^{-1}$. In every case laser power on the sample was always less than 1 mW. [31]

Near normal infrared spectra were taken on heating using two experimental facilities, one corresponds to reflection and a second one to emission. From 12 K to room temperature and from 300 K up to about ~850K we have measured reflectivity at 1 cm$^{-1}$ resolution with two Fourier transform infrared spectrometers, a Bruker 66 V/s, and a Bruker 113V, respectively, with conventional near normal incidence geometries. Low temperature runs were made with the sample mounted in the cold finger of a Displex closed cycle He refrigerator. For high temperature reflectivity a heating plate adapted to the near normal reflectivity attachment was used in the Bruker 113V vacuum chamber. For the low temperature set an evaporated in situ gold film was used as 100% reference reflectivity while a plain gold mirror was used for reflectivity between 300 K and 800 K.

Infrared Emissivity in the ~500 K to ~1700 K range at 2 cm$^{-1}$ resolution was measured with two Fourier transform infrared spectrometers, Bruker Vertex 80v and Bruker Vertex 70, coupled to a rotating table placed inside a dry air box allowing to simultaneously measure the spectral emittance in two dissimilar spectral ranges from 40 cm$^{-1}$ to 13000 cm$^{-1}$. The sample, which was heated with a 500 W pulse Coherent $CO_2$ laser, was positioned on the rotating table at the focal point of both spectrometers in a position equivalent to that of the internal radiation sources inside the spectrometers. In this measuring configuration, the sample, placed outside the spectrometer, is the infrared radiation source, and conversely, the sample chamber inside the spectrometers is empty.[32,33,34]

Detail extended account on the experimental facilities used in collecting data and their analyses for the research reported here may be found in the Supplemental Material.[28]

## METHODS AND DATA ANALYSIS

We retrieve the normal spectral emissivity, $E(\omega, T)$,

$$E(\omega,T) = \frac{FT(I_S - I_{RT})}{FT(I_{BB} - I_{RT})} \times \frac{\mathscr{P}(T_{BB}) - \mathscr{P}(T_{RT})}{\mathscr{P}(T_S) - \mathscr{P}(T_{RT})} E_{BB} \qquad (1)$$

after measuring three $I$ interferograms, i.e., sample, $I_S$; black body, $I_{BB}$; and, environment, $I_{RT}$, and where $FT$ stands for Fourier Transform. $\mathscr{P}$ is the Planck's function taken at different temperatures T; i.e., sample, $T_S$; blackbody, $T_{BB}$; and surroundings, $T_{RT}$. $E_{BB}$ is a correction that corresponds to the normal spectral emissivity of the black body reference (a LaCrO$_3$ Pyrox PY 8 commercial oven) and takes into account its non-ideality.[32]

The temperatures at which the measurements were done were calculated using the Christiansen point, i.e., the frequency where in insulators the refraction index is equal to 1, and the extinction coefficient is negligible. At that frequency, the emissivity $E(\omega, T)$, eq. (1), is set equal to 1, and since Christiansen point varies its position with temperature, after identifying the frequency for which $E(\omega_{Christ}, T_{sample}) = 1$ the only remaining variable left is T$_{sample}$, the sample temperature that univocally corresponds to it. This procedure becomes a less reliable temperature reference when hopping carriers conceal or distort the frequency minimum.

After acquiring the emissivity data, we placed our spectra in a more familiar near normal reflectivity framework using the second Kirchhoff law, that is,

$$R = 1 - E \qquad (2)$$

where $R$ is the sample reflectivity.

Then, knowing that

$$R(\omega) = \left| \frac{\sqrt{\varepsilon^*(\omega)} - 1}{\sqrt{\varepsilon^*(\omega)} + 1} \right|^2 \qquad (3)$$

we calculate the dielectric function from the measured reflectivity or emissivity using a multioscillator fit [35] given by,[36]

$$\varepsilon(\omega) = \varepsilon_1(\omega) - i\varepsilon_2(\omega) = \varepsilon_\infty \prod_j \frac{(\omega_{jLO}^2 - \omega^2 + i\gamma_{jLO}\omega)}{(\omega_{jTO}^2 - \omega^2 + i\gamma_{jTO}\omega)} \quad (4)$$

at every temperature. $\varepsilon_1(\omega)$ is the real and $\varepsilon_2(\omega)$ the imaginary part of the dielectric function (complex permittivity, $\varepsilon^*(\omega)$); $\varepsilon_\infty$ is the high frequency dielectric constant taking into account electronic contributions; $\omega_{jTO}$ and $\omega_{jLO}$, are the transverse and longitudinal optical mode frequencies and $\gamma_{jTO}$ and $\gamma_{jLO}$ are their respective damping.[37] The dielectric simulation parameters for $h$-ErMnO$_3$ showing TO-LO phonons and a broad mid-infrared oscillator band between ~1000 cm$^{-1}$ and ~10000 cm$^{-1}$, assigned below to polarons, are shown in tables in the Supplementary Material.[28] Note that **at** higher temperatures, an extra high intensity narrower band corresponding to a d-d transition, has been added centered at ~12000 cm$^{-1}$ and that for the ~1624 K spectrum we also added a Drude term in the analysis.[28]

Next, with the knowledge of individual contributions to $\varepsilon_2(\omega)$, we calculate the real part of the temperature dependent optical conductivity, $\sigma_1(\omega)$,[38] given by

$$\sigma_1(\omega) = \frac{\omega \cdot \varepsilon_2(\omega)}{4\pi} \quad (5)$$

This constitute our experimentally measured optical conductivity. Tables I and II show in brackets the phonon reststrahlen mid-value and MIR band ($2*E_{bipolaron}$) peak frequencies from the measured reflectivities and emissivities shown in figures and in tables in the Supplemental Material section.[28] They then allow a quantitative matching in the MIR to the polaron formalism discussed in Sec. V (Figs. **8 and 9).**

## RESULTS AND DISCUSSION

### A. *Low-frequency phonons, magnetic excitations, and Er$^{3+}$ crystal field transitions in the ferroelectric phase*

Fig. 2(a) shows the phonon profiles in the lower temperature regime from 12 K to 300 K where reflectivity spectra and multioscillator fits (tables SII- SIV) in the Supplemental Material[28] allow us to identify 21 of the 23 vibrational bands predicted by the irreducible representation for the hexagonal space $P6_3cm(C^3_{6v})$ Z=6 (**Ref. 39**),

$\Gamma_{IR}= 9A_1+ 14E_1$                             eq. (6)

We also corroborate that they undergo an overall regular hardening and band narrowing in close similarity with isomorphous YMnO$_3$ (**Refs. 40, 41**) and HoMnO$_3$(**Ref.42**). Lattice phonons around 200 cm$^{-1}$ undergo relative intensity changes as the Rare Earth in two inequivalent sites reaccommodate in the ferroelectric phase. Between ~292 cm$^{-1}$ and ~340 cm$^{-1}$ there is a set of internal modes from which one has a distinctive reststrahlen screening that will be addressed in the following sections. The strongest of these modes (Fig 2(c)) gradually split as the temperature get close to the antiferromagnetic onset near T$_N$~79 K due to spin-phonon interactions[17] and tripling unit cell. Overall, our spectra, that being from ceramics have A$_1$ and E$_1$ phonons superposed, are also in agreement with lower temperature E$_1$ transverse optical (TO) modes ($E \perp c$) and A$_1$ listing by Basistyy et al.[17]

A more complex environment is suggested by lower frequency near normal reflectivity revealing a well-defined tail of a broad band at THz frequencies (Fig. 2,(b)). We understand the band, THz centered, as arising from structural inhomogeneities, unreleased strains, and potential freer charges localized in grain boundaries or ferroelectric domain walls. The electrical response of the interlocked antiphase boundaries and ferroelectric domains walls is dominated by bound charge oscillations.[43] Vacancies and interstitials may also act as local scattering centers that relate to some structural disorder at atomic level. Local structural inhomogeneities and an imperfect unit cell tripling recreate a relaxor type scenario that may be regarded as a remnant of an incomplete order-disorder phase transition subjected to the ordered polar regions in **c** axial ferroelectric $h$-ErMnO$_3$. This heterogeneity is generically known as the origin of the so-called boson peak[44] that is well reproduced in the interval 12 K and 300 K by fitting to a Gaussian profile (Tables SII-SIV).

We have also addressed this spectral region in more detail in our BESSY II absorption low alpha runs.[28] Between 2.4 K and 5 K the low temperature magnetic phase diagram of $h$-ErMnO$_3$ shows field dependent increasing coupling between the $Er^{3+}$ and the out of plane $Mn^{3+}$ moments suggesting a complex 3d and 4f interplay. In this temperature range $Er^{3+}$ moments at the Wyckoff 2a site align ferromagnetically along the *c* axis inducing spin reorientation at $Er^{3+}$ 4b´s and Mn sites. This means going from the $\Gamma_4$ into the $\Gamma_2$ representation at temperatures lower than $T_N$.[14,45] Under the umbrella of the Gaussian bell-shaped reflectivity we detect two features, $M_1$ and $M_2$, that harden on applied fields that, associated to magnetic excitations are in agreement with earlier assignments by Chaix et al.[18] Fig. 3 (a) and 3(b) shows the relative effect of applied fields up to 7 T on the $M_1$ magnon-like excitation as well as the behavior for $M_2$. In this last case, the band is quite broad with a profile that may be thought as made of several components rather weak and disperse. When increasing fields ferromagnetism is induced,[14] the two band envelop for 2T (Fig. 3(d)) suggests spin precessions due to an increment in related anisotropies. Higher fields up to 4 T further increase the magnetic distortion as if reorienting some of them, after which, being overwhelmed by the field coalesce into one net precessing mode at 5 T with a prevailing spin relative to the spatial direction also found for $M_1$ mode (Fig. 3(c)). This suggest a composed scenario by which magnetic moments at rare earth sites in ferroelectric domain walls create domain wall magnetism from uncompensated $Er^{3+}$ moments[46] in an environment of fluctuating $Mn^{3+}$ antiferromagnetic order.[47]

At slightly higher frequencies, there are crystal field transitions from Kramers ion $Er^{3+}$ (doublet $4f^{11}$)(**Ref.48**) echoing the two non-equivalent lattice sites. The relative changes induced by the applied magnetic field to the zero- field cooled spectrum on the 0.7 T isoline are shown in Fig. 4(a). The crystal field levels of $Er^{3+}$ and several much weaker features are clearly delineated as the samples passes from being ferrimagnetic to ferromagnetic induced. Our data is in full agreement with Chaix et al having only a minor energy difference in the $E_2$ line that resolves when their neutron data is properly deconvoluted.[18] In addition, our spectra point to a remarkable and unexpected correlation between the two most intense of these lines and a couple of higher frequency weaker but sharp peaks. Dotted lines in Fig. 4 show that absorption of the main crystal field transition of antiferromagnetically coupled $Er^{3+}$(4b) at 62.0 $cm^{-1}$, Fig.4 (b), match a sharper counterpart at exactly the double energy of 123.9 $cm^{-1}$ in either antiferromagnetic (1T) or ferromagnetic phases (4T, 7T) at 5 K. The same kind of correlation, now shown as minima in field depletion perhaps reflecting the more limited coupling, is observed for $Er^{3+}$(2a) at ~82.9 $cm^{-1}$ band against 165.8 $cm^{-1}$ respectively. Each pair corresponds to each of the two $Er^{3+}$ ions at the non-equivalent sites. There is also an undefined background continuum that goes unmatched but may be related to interfaces with magnetic and/or electric order parameters moving at unison with the applied field when ferromagnetism is induced in the sample.[14]

We also find that those replicas have a remarkable match in our Raman and reflectivity spectra (see Fig.4 (c)) at the same frequencies where two weak and narrow but well defined phonon bands may be traced up to lock-in Tc ~830 K, merging into the high temperature background. They also agree with phonon profiles in isomorph $YMnO_3$ with $Y^{3+}$ non-magnetic.[40] That is, by calculating applied to zero field ratios and comparing with reflectivity and Raman spectra[31] we put in evidence a pure magnetic and phonon electric dipole counterpart of a unique multiferroic excitation. While their specific origin still remains a puzzle (being detected with three independent set ups and techniques rules out the possibility of an interferometer or detector artifact) we may reason of these are due to the strong anisotropies between the Rare Earth and oxygens involved in plane buckling and bipyramid tilting in conjunction with the dynamics emerging from the *c* axis negative thermal expansion[49] and, perhaps, also associated to reinforcing ferroelectricity being the $4f^{11}$-$3d^4$ interplay a known player in polar ordering[45] in an on-going process with imperfect (√3x√3)R30° structural ordering.[50] It points to a symmetry allowed lattice phonon interactions with 4f electrons associated to ground $Er^{3+}$ crystal field levels that is better seen at low temperatures due to the increase in their population. Very recently, Singh et al[51] reported that in double perovskite $Nd_2ZnIrO_6$ low lying excitations amount to strong coupling between phonons and $Nd^{3+}$ $4f^3$ crystal field excitations raising the possibility that all these magnetic and lattice degrees of freedom have common grounds that might also be shared by features found in $HoMnO_3$ **(Ref. 52)**

B. *Screening of an Optical Phonon Reststrahlen*

Oxides have a strong tendency toward delocalized p Oxygen states in which the interplay of Coulomb and electron-phonon interaction lead to a ferroelectric instability.[53] It is known that the origin of the anisotropies in the ion polarizability lies in the $O^2$ instability as free ion.[54, 55] It means that in the *h*-$ErMnO_3$ lattice forming a closed 2p6 shell Oxygen needs neighboring Mn and Rare Earth. Within this context, as it is shown in Fig. 6**,** most lattice and bipyramid internal modes continue above room temperature the softening and thermal broadening trend already shown between 12 K and 300 K (Fig. 2(a)). The distinctive exception is a narrower and strong asymmetric profile that between 292cm$^{-1}$ and 333 cm$^{-1}$ (Table SII) that has a lower intensity step-like plateau on the higher frequency side. This is the vibrational region with $A_1$ and $E_1$ internal modes mainly involving in- and out- of plane oxygen motions,[16,39,41,42] that through hybridization add to the net polar distortion. The shoulder remains partially screened approaching room temperature in a split mode picture (Fig. 2(c)) while decreases approaching ~830 K to become almost fully screened at ferroelectric lock-in Tc (arrow in Fig. 5). The result is an asymmetric profile that at that may still be fitted with a single oscillator (Table SVII).[28] In the intermediate phase, between ~830 K and ~1500 K, Figs. 5 and 6(a), it becomes barely distinctive from a horizontal trace pointing to a strong anomaly in the reststrahlen macroscopic field associated zone center ~273 cm$^{-1}$ longitudinal optical mode (1119 K, table SVIII).[28]

We recall that a reststrahlen macroscopic field arises from the long-range Coulomb interaction responsible of the transverse optical-longitudinal optical (TO-LO) split. It may be though as if the LO mode were backed by long range electric fields appearing as an extra restoring force due to a Coulomb interaction. The net result is the increase of the longitudinal optical mode frequency. Implies a relative displacement of charged ions in the lattice unit cell generating a macroscopic field that can then be expected to interact with electrons. The interaction between an electron charge and the macroscopic Coulomb potential is known as Fröhlich interaction.[56]

It is also known that carrier mobility in a polar lattice is reduced by coupling to LO modes via Fröhlich interaction in what may be basically understood as due to a polar lattice-charge carrier electrostatic interaction leading by localization to the formation of small polarons. Thus, when the macroscopic field gets electron screened the TO-LO split becomes narrower reducing the phonon oscillator strength as observed. Figure 6(a), and supplemental Figs. S11 and S12, show that close to the paraelectric phase transition, up to ~1500 K, there is already no trace of this phonon. At these temperatures the main change corresponds to lattice modes around ~190 cm$^{-1}$ gradually merging into one symmetric broad band denoting a sole Rare Earth crystallographic position onset of the paraelectric phase with a net inversion center not allowing off-center displacements.

## C. *Fröhlich Raman Resonance*

As $h$-ErMnO$_3$ at room temperature is a non-centrosymmetric polar compound the infrared active modes are also Raman zone center allowed implying that we may further pursue the characterization of the anomalous reststrahlen screening by measuring its Raman activity. It is known from earlier studies on hexagonal semiconductors that charge density fluctuations of free electrons couple to the macroscopic longitudinal electric field of infrared active polar optical lattice vibrations and that the resulting coupled modes have a mixed plasmon-phonon character.[22] Phonon-plasmon coupling was intensely studied optimizing cross section by choosing exciting laser lines with energies above the semiconducting gap, in which case, the zone center LO mode cross section is strongly enhanced via Fröhlich interaction, again, the interaction between an electron charge and the macroscopic Coulomb potential.[22]

First order Raman active zone center modes of $h$-ErMnO$_3$ belong to the irreducible representation[39]

$$\Gamma_{Raman}= 9A_1+14E_1+15E_2 \qquad \qquad eq\ (9)$$

These Raman phonon profiles are heavily dependent on the excitation line and have as dominant feature the bipyramid breathing and its overtone. This $A_1$ 683 cm$^{-1}$ mode ($\lambda_{exc}$=457 nm, 300 K ), Fig. 7 (a), involves $c$ axis apical O1-O2 ions displacements creating an effective polar distortion in the MnO$_5$ bipyramids along interplanar ferroelectric $c$ direction.[42] It is associated to the transition between xy(x$^2$-y$^2$) and 3z$^2$-r$^2$ orbitals by which the occupation of the 3z$^2$-r$^2$ orbital creates a Coulomb repulsion enhancing the breathing relative to the Mn$^{3+}$ ions in the plane.[16] Our room temperature spectra (Fig. 7 (a)) are in overall agreement with earlier Raman data.[31]

Here we will limit our comments to the spectral range specific to the region in which we observed the unusual changes in the infrared spectra commented in the preceding paragraphs.[31] Specifically, we use 355 nm laser exciting line aiming to avoid the interference of electronic transitions in the cross sections. It allows the detection of a strong unambiguous quasi-resonant band (Fig. 7(b)) emerging near the middle frequency of the reststrahlen band down from the LO frequency that may be straightforwardly associate to the carrier screening. TO modes in the ferroelectric phase are both Raman and infrared active while the LO's are Raman active only. The peaking at 322 cm$^{-1}$ is at shoulder mid-frequency of the infrared reflectivity where the mode at 297 cm$^{-1}$ (TO) may be used as reference to confirm the agreement between Raman and infrared spectra (Fig. (7(b), inset)). The LO resonance is red shifted relative to the infrared minimum because the presence of carriers screening of the macroscopic field, Landau damping, in the polar environment suggesting the association with conclusions by Abstreiter et al on n-type, p- type Si, Ge, and n-type GaAs semiconductors.[22, 57, 58, 59] It is however worthwhile to note that our band screening is due to delocalized carriers subjected to the dynamics of the negative thermal expansion in an environment prone to strong polaron coupling and weaker deformation potential mechanisms involving intraband excitations of carriers as shown in the sections.

## D. *Negative c axis thermal expansion*

A dynamical perturbative mechanism that requires attention in our analysis is in the uniaxial negative thermal expansion reported in the $h$ RMnO$_3$ (R=Rare Earth, Y) family of compounds.[50, 60, 61, 62, 63, 64]

The relative negative lattice distortion along the *c* axis, Fig.1(b), is stronger in the intermediate phase, where the structural trimerization involves Mn and O ion displacements through plane buckling due to mismatched Er ions. At about the lock-in temperature it turns gradually into a more lower temperature moderate regime.[50, 60]

Negative thermal expansion is understood as due to rigid polyhedral lattice units performing a quasi-rigid rotation (libration).[65,66] These polyhedras in $h$-ErMnO$_3$ are MnO$_5$ bipyramids in which the corner sharing bridging oxygen in M-O-M links may be seen pivoting as a hinge dynamically increasing the oxygen amplitude. This by purely geometric arguments reduces the X-ray measured bond length of the true lattice constant, since now, it appears as a shorter distance when the bond is subjected to positive and negative strong fluctuations.[67, 68] For $h$-ErMnO$_3$ we associate this motion to vibrational modes centered around 290-370 cm$^{-1}$ involving upward motions of the O ions (and Mn) in the stiffer bipyramid sublattice.[16, 62] Band broadening and softening anharmonicity in other phonons mostly contribute to positive thermal expansion but sharper band changes around the ~370 cm$^{-1}$ internal mode region, near and below T$_C$, also indicate processes for which the bipyramids may be better though of as quasi-rigid units. The out-of-plane displacements add to ferroelectricity[26] considering Oxygen ions polarizable objects.[53, 54, 55]

In a more quantitative picture, the gradual evolution of the screening in the ~322 cm$^{-1}$ TO-LO macroscopic field split would be then an indication of a disruption in the strong electron-electron correlations in *h*-ErMnO$_3$. Local reststrahlen screening would be tied to the sublattice motion singular to negative thermal expansion. Below ∼830 K the negative thermal expansion along *c* is gradually allows electrons to add coherently to the polar mechanism of the already present non-centrosymmetric Rare Earth displacements that is at the structural origin of ferroelectricity below at ∼1500 K.

This scenario connects our compound to earlier ideas on the possibility of a Mott semiconductor[69] having a ferroelectric instability[70] and assimilate this to dynamically disrupted local electron-electron interactions of Mott charge carriers inducing Coulomb screening in the reststrahlen TO-LO split. Those carriers gradually coalesce condensing into collective charge localization and ordering locking-in ferroelectric electrons below $T_C$ ∼830 K. Basically, this may be then understood as a locking-in electronic process for a macroscopic spontaneous polarization response of highly correlated electrons.[21] However, our spectra show that band sharpening of the internal modes is not limited to the screened mode but rather extend to the neighboring vibrational modes that emerge as sharper bands (Fig.5) below Tc from the contour that it is already outlined in the intermediate higher temperature phase (Figs. 5 and 6). This is a known activity for phonons in ferroelectric insulators sustaining an incommensurate phase and tripling of the unit cell.[71] For this reason, and although this behavior compounds with the dynamics imposed by the negative thermal expansion, we propose that $T_C$ ∼830 K in *h*-ErMnO$_3$ be considered as a lock-in Curie temperature. Ferroelectric coherent order is reduced on heating above Tc~830 K at the change in the slope of the uniaxial negative thermal expansion in what amounts a *c* dynamic contraction of Oxygen-Erbium bonds at a temperature range in which, as described in the next section, there is a change on cooling from a bipolaron to a more localized small polaron regime. It is also appealing to think that the anomalies observed may relate to local bond conductivity increments in an environment for which low disorder increments break down an initial Mott insulating system **as** recently predicted for interacting systems.[72]

## E. *Small polaron and higher temperature bipolarons*

We have seen in previous sections the main role of electron-lattice correlations that may also be functional to the magnetic order. This merits the extension of our quantitative look to the mid-infrared region (MIR) where electron-phonon interactions in the ferroelectric environment are key for revealing quasiparticles named small polarons.[73]

A small polaron is made of an electron laying in the ion potential created by the disturbance that the charge carrier creates. Short range electron-phonon interactions are paramount.[74]

Its optical detection is due to a self-trapped carrier excited from its localized state to a localized state at a site adjacent to the small polaron site. The small polaron range is usually less than the unit cell size having as main transport property a characteristic thermal activated hopping accompanied by the lattice deformation. The implied picture is found in systems than range from oxides to polymers and has been the subject of many reviews.[24,25,75]

As already pointed in the Supplemental Material Methods and Data Analysis section,[28] our reflectivity (or 1-emissivity for measured emission) multioscillator fits yield individual contributions to the imaginary part of the dielectric function, $\varepsilon_2(\omega)$, that in turn, allows using eq(5) to calculate the real part of the temperature dependent optical conductivity, $\sigma_1(\omega)$, spanning from the far- to the mid-infrared. As it was stated, this constitute our data from which the mid-infrared will be examined in terms polaron quasiparticles.

Our starting point for matching the experimental optical conductivity is the known formulation by Reik and Heese for small polarons.[23,76] Here, small polarons are addressed microscopically as due to nondiagonal phonon transitions. The optical conductivity is calculated for carriers in one small band and interband transitions are excluded. Starting with a Holstein´s Hamiltonian,[77] the frequency dependent conductivity is calculated using Kubo´s formula.[78,79] The real part of the model optical conductivity for a finite temperature T, $\sigma_1(\omega,\beta)$ is given by

$$\sigma_1(\omega,\beta) = \sigma_{DC} \frac{\sinh\left(\frac{1}{2}\hbar\omega\beta\right)\exp\left[-\omega^2\tau^2 r(\omega)\right]}{\frac{1}{2}\hbar\omega\beta\left[1+(\omega\tau\Delta)^2\right]^{1/4}}, \quad (10)$$

$$r(\omega) = \left(\frac{2}{\omega\tau\Delta}\right)\ln\left\{\omega\tau\Delta+\left[1+(\omega\tau\Delta)^2\right]^{1/2}\right\} - \left[\frac{2}{(\omega\tau\Delta)^2}\right]\left\{\left[1+(\omega\tau\Delta)^2\right]^{1/2} - 1\right\}, \quad (11)$$

$$\text{with } \Delta = 2\varpi_j \Psi \tag{12}$$

and the relaxation time

$$\tau^2 = \frac{\left[\sinh\left(\frac{1}{2}\hbar\varpi\beta\right)\right]}{2\varpi^2 \eta} \tag{13}$$

That is, our model conductivity, $\sigma_1(\omega,\beta)$, $\beta=1/kT$, is a bell shaped quasi-Gaussian three parameter dependent; $\sigma_{DC}=\sigma(0,\beta)$, the nominal electrical DC conductivity; the frequency $\varpi_j$ that corresponds to the average between the transverse and the longitudinal optical mode of the $j^{th}$ restrahlen band; and $\eta$, is a parameter characterizing the strength of the electron-phonon interaction, i.e., the average number of virtual phonons contributing to the polarization around a localized polaron. These are in eq.(10) the adjusting parameters to the MIR for experimental conductivity extracted in the analysis of the reflectivity (or 1- emissivity for measured emission). Allows to identify the phonon frequency, $\varpi_j$, associated to the polaron from those out of the listing modes shown in fit tables in reference **28**. $\eta_j$, that constitutes the only true free parameter, is also proportional to the small polaron binding energy $E_b=\eta\varpi_j/2$. The other constant, $\sigma_{DC}=\sigma(0,\beta)$, is just multiplier factor.[80] $\eta\sim3$ implies a weak electron-phonon interaction while a value around 14 or higher would correspond to the very strong end.[81]

To reproduce the full mid-infrared spectral region, we then allowed the possibility of more than one vibrational contribution using an empirical approach shown successful when applied to other distorted perovskites and glassy systems.[82, 83, 84] This results in the convoluted addition of bell shaped uncorrelated individual contributions in the region from 1000 cm$^{-1}$ to 9000 cm$^{-1}$, each calculated at a phonon frequency $\varpi_j$ associated to the strength $\eta_j$ at a temperature T.

A good fit of the experimental optical conductivity of *h*-ErMnO$_3$ at 300 K requires a full small polaron approach with torsional (319 cm$^{-1}$), breathing (735 cm$^{-1}$), and overtone (1480 cm$^{-1}$) as distinctive vibrational frequencies (Fig. 8, Table I) for each quasi-Gaussian. Discussed in previous sections, these peak positions correspond to the band with the anomalous reststrahlen and the interplane ferroelectric mode (and its overtone) respectively. The same picture holds down to about 100 K where reflectivity (Fig. 2 (c)) shows the unscreened portion of the band profile from 290 cm$^{-1}$ to 330 cm$^{-1}$ starts splitting due to spin-phonon interactions[17] and the on-going unit cell tripling. At 12 K only invoking independent vibrational group modes satisfy the mid-infrared response shown in Fig 8 that is also seen in overall sharper defined reflectivity (Fig. 2). All parameters used in the computation of these lower temperature optical conductivities are collected in Table I. Below each frequency $\varpi_j$ we also added in brackets the mid TO-LO split measured frequency, from the reflectivity fit, shown in tables for 300 K, 109 K, and 12 K of the Supplemental Material.[28] In every case, our $\eta_j$´s , the parameter characterizing electron-phonon interactions, are found in the strong regime concomitant with the Raman measurements discussed in Sec. IV C on the Fröhlich resonance.

On the other hand, increasing temperature, yields MIR profiles that cannot be totally satisfied by only the quasi-Gaussian small polaron contributions under consideration. At 545 K, (Fig. 8) still below the ferroelectric lock-in temperature, we found that in addition of a small polaron quasi-Gaussian with the ~707 cm$^{-1}$ breathing vibrational mode (Table I) it was necessary to add a bipolaron term (Table II) for reproducing the lower frequency portion of the optical conductivity, meaning, that intermediate temperatures conductivities reflect different degrees of localization and coexistence in the a changing stripe/vortex environment. [26]

A bipolaron is made of two small polarons in association as two carrier self-trapped localized bound pair, in which the common potential exceeds their Coulomb repulsion.[84] In this regime the confinement energy of small bipolarons is double relative to that in small polarons because the presence of the second electron, i.e., the depth of the well that self-traps both carriers is twice as deep. Their effective mass is expected to be also much larger than the hundred-fold increments of the electron mass expected for small polarons.[24, 74, 85, 86]

The electron-phonon coupling energies are quadrupled, and at the same time, both charged carriers repel through their U- Coulomb interaction ,i.e., the bipolaron is stable if $2E_b>U$. Accordingly, band peaking associated to bipolarons is found at higher frequencies than for small polarons.[87] This means that for small bipolarons the $2E_b$, the peak energy of the small polaron absorption, has to be replaced by $2E_b'=4E_b-U$, the frequency of the measured absorption (conductivity) maximum, and $\Delta$ by $\Delta'= \sqrt{2E_b'E_{vib}}$, where $E_{vib}$ corresponds to a relevant mediating phonon.[24] Using this simplified version for the optical conductivity, that supplies the absence of the microscopic approach but retains the basic concept of Reik's small polaron original proposition, reads

$$\sigma(\omega, \text{T}) = \sigma_{DC} \frac{sinh(4E_b'\eta\omega/\Delta'^2)}{4E_b'\eta\omega/\Delta'^2} \exp(-(\eta\omega)^2/\Delta^2) \quad (14)$$

where $\sigma_{DC}=\sigma(0,\beta)$, $\beta=1/kT$, is again, the electrical or optical DC conductivity.

We found that this formulation reproduces the MIR optical conductivity between the Christiansen point and the absorption edge of the Mn $d_{x2-y2,xy} \rightarrow d_{3z2-r2}$ transition in our highest temperature regime of the emission measurements. Our bipolaron band profiles are similar to the ones earlier reported for other oxides.[88, 89] However, unlike in distorted perovskites, where a successful fit requires the octahedral breathing frequency ~700 cm$^{-1}$ for the bipolaron mediating phonon $\omega_{(vib)}$, in $h$-ErMnO$_3$ we need to replace that value for a frequency close to that anomalous band at ~322 cm$^{-1}$. This brings the quasi-Gaussian (eq. 14) to agreement with the optical conductivity from measured 1-emissivity. Figure 9 shows these at different temperatures in the intermediate region between ~830 K and ~1500 K and in the paraelectric phase at 1545 K. 2$E_b$' from model fit and 1-emissivity (in brackets) are listed for each temperature in Table II. It can be seen that the bipolaronic band hardens between ~700 K and ~1200 K. Further up in temperature the opacity in the compound increases flattening the profile, which starting at ~1300 K, is seen as prompting a discontinuity toward the paraelectric phase. The MIR evolution reflects the stripe from orderly aligned stripes to the curly, close to $T_{INC}$, to the articulated vortex-antivortex domain structure pattern reported in the paraelectric phase for $h$-ErMnO$_3$.[26]

At 1578 K (Fig. S9)[28] our optical conductivity fit (Fig. 9) yields a frequency at ~415 cm$^{-1}$ to reproduce the experimental data. This frequency matches the average of the three torsional bipyramid internal modes (asterisks in Table SX)[28] that have as common denominator the Oxygen ion displacements, these being, part of the five detected out of six zone center modes, $\Gamma_{IR}=3A_{2u}+3E_{1u}$, predicted by group theory for the space group P6$_3$/mmc(D$^4_{6h}$) in the paraelectric centrosymmetric phase.[39]

Increasing the temperature further Fig. S10 shows that electron hopping induces a lower frequency Drude term (eq. S5) onset of an insulator-metal phase transition at ~1600 K in a sequence that at still much higher temperatures leads to sample decomposition. This effect, as well as possible embedded temperature induced defects,[82] increases conductivity prompting an insulator into an incipient poor conducting solid yielding a quantitative picture for a thermally driven insulator-metal phase transition. The new tail in the far infrared, as pointed out Sec. II, also reduces the usefulness of the Christiansen point as functional thermometer.[28]

Above 1600 K, Fig 9, the bipolaron peak position softens yielding a mid-infrared spectrum that beyond ~3000 cm$^{-1}$ is an undistorted tail up to 10000 cm$^{-1}$ similar to the a mid-infrared response for single small polarons (Fig. 8) suggesting localization in an intermediate regime for a many body scenario. In this scenario one of the dimerized electron pair constituting the bipolaron hopes to a next near neighbor small polaron site changing the mid-infrared profile within picture that turns indistinguishable from what we already reported for the high temperature metal-insulator transition in orthorhombic O-ErMnO$_3$ at about the same temperatures.[89] I.e., as shown in Fig. 9 the reproduction of the optical conductivity at ~1624 K is a plain superposition of contributions product of the bipolaron dissociation as for small polarons coexisting with Drude thermal activated itinerant carriers. The small polaron parameters used in this fit are shown in Table I while those for the bipolaron band are in Table II.

## Conclusions

In an effort to get a more comprehensive picture on multiferroic ErMnO$_3$ in its hexagonal phase we studied the infrared response using absorption, reflectivity, and emissivity techniques in its full range of existence. An intriguing phonon magnetic convergence was found at the lowest frequency modes. Phonons, whose detection extends up to the onset of the intermediate phase, correspond to well defined magnetic replicas of Er$^{3+}$ crystal field transitions suggesting strong magnetoelectric couplings.

Our measurements show that in addition to the known ion displacement in MnO$_5$ bipyramid buckling, and Rare Earth vertical shift away from the high temperature mirror plane,[9, 26] there is a dynamical bond instability related to ferroelectric order. Its origin is potentiated by the oxygen intrinsic polarizability[54, 55, 90] bringing up a critical interplay of ionic and electronic interactions to a common ground with that reported for many ferroelectrics.[73]

A most distinctive temperature dependent phonon profile is found associated with a torsional mode involving oxygen displacements dynamically correlated to the negative thermal expansion along the *c* axis.[50,60] It has a nearly fully screened macroscopic field in the intermediate phase where it is also found mediating in bipolaron conception. Below the lock-in ferroelectric transition at ~830 K it turns into a well-defined asymmetric band, at 288 cm$^{-1}$-329 cm$^{-1}$ (300 K), that having one remaining component partially screened further splits as spin-phonon interactions become significant and the unit cell tripling takes place. Its existence implies an environment revealing a delicate balance of long- and short-range interactions that compounds with local disruptions of the strong Coulomb electron repulsion in the non-centrosymmetric entangled phase. Concomitant with this, but using Raman scattering excited with $\lambda_{exc}$ 355 nm above the gap laser line, we verified strong electron-phonon interactions via a Fröhlich resonance at 322 cm$^{-1}$ induced by carrier delocalization interplay with the Coulomb originated TO-LO split of the corresponding screened reststrahlen. This phonon also remains a fundamental feature when the high temperature bipolaron regime turns to small polaron. At these temperatures, the MIR optical conductivity is reproduced using Reik's formulation.[23] Accordingly, we associate that vibration with a primary order parameter.

At 1578 K, in the paraelectric phase, we found that a fit using a vibrational frequency at ~415 cm$^{-1}$ results in an almost perfect matching with the experimental optical conductivity. This frequency corresponds to the average of three main torsional bipyramid internal modes (Table SX) involving oxygen displacements. High-temperature MIR profiles change as the sample opacity increases when straight stripes turn curly toward a more complex domain figures above ~1500 K yield vortex-antivortex pattern.[26] Single small polarons and bipolarons coexists at higher temperatures where a Drude term is found marking an insulator-metal phase transition at ~1600 K. This electron hopping onset signals sample decomposition at still much higher temperatures.

We conclude that knowledge of the full temperature dependence of the phonon spectrum now allows to draw similarities between *h*-ErMnO$_3$ and improper ferroelectrics that sustaining an intermediate phase become lock-in ferroelectric in a commensurate low temperature superstructure tripling the paraelectric unit cell.[91] Inside this broader framework, pseudohexagonal K$_2$SeO$_4$, and isomorphs X$_2$ZnCl$_4$ (X=K, Rb), with a displaced and order-disorder paraelectric-incommensurate phase transition, respectively, are known prototypes. The delicate balance of long- and short-range interactions for the second order soft mode found at T$_{INC=}$129 K in K$_2$SeO$_4$,[92] is lost to a first order order-disorder transition by isotopic replacement as in isomorph Rb$_2$ZnCl$_4$ (Ref [93]), a dynamics that seems to be shared in the paraelectric to non-centrosymmetric transition of h-ErMnO$_3$.

In a phase that it is lattice incommensurate both, long wavelength polarization and strain wave, coexist[27] and there are interactions that tend to restore periodicity growing domains with commensurate structure and spontaneous polarizations of opposite signs separated by discommensuration regions (it conforms a soliton regime that involves domain patterns)[94] a fact compatible with reports on polar findings in the *h*-ErMnO$_3$ intermediate phase. It is also concurrent with quasielastic X-ray scattering lines of frozen modes due to imperfect ($\sqrt{3}$x$\sqrt{3}$)R30° ordering in *h*-ErMnO$_3$ (Refs. **50, 60, 95**) in a picture that compounds with the dynamics imposed by the negative thermal expansion and the high temperature.

As the sample cools the soliton density diminishes growing the coherence of fluctuations that condensing triggers the ferroelectric phase at T-Curie-lock-in. The result is a net macroscopic ferroelectric polarization along *c*. Then, the intermediate phase in *h*-ErMnO$_3$ may be though as the result of lattice instability against the creation of solitons[96,97] in a scenario for discommensurations (domain walls)[98] with latent hidden lattice symmetry. [95,99-101] Accordingly, we suggest naming in *h*-ErMnO$_3$, T$_{INC}$~1500 K as the onset temperature for the intermediate phase that ought to be though incommensurate while the transition at Curie T$_C$ ~ 830 K would signal the ferroelectric phase locking-in.

# Acknowledgments

NEM is indebted to the laboratory on Conditions Extrêmes et Matériaux: Haute Température et Irradiation - UPR3079 CNRS (C.E.M.H.T.I.)) and staff in Orléans, for sharing expertise on research and financial support performing far infrared reflectivity and emissivity measurements. He also thanks BESSYII at the Helmholtz-Zentrum Berlin für Materialien und Energie for beamtime allocation under proposal 181-06471-ST-1.1-P and 172-05636-ST/R-1.1-P and for financial assistance supported by the project CALIPSOplus under the Grant Agreement 730872 from the European Union Framework Program for Research and Innovation HORIZON 2020. JAA acknowledges the ILL-Grenoble for the allowed neutron time, and the financial support of the Spanish "Ministerio de Economia y Competitividad" (MINECO) through Project Nº MAT2017-84496-R.

# TABLE I

Parameters of small polaron fits to the *h*-ErMnO$_3$ experimental optical conductivity at 545 K, 300 K, 105 K and 12 K (see text). Note that resulting vibrational frequencies $\varpi_{phj}$ (j=1,2,3) are in agreement with the experimental reststrahlen mid-value frequencies shown in brackets. The DC conductivity, $\sigma_{DC}$, in eq(10) was kept as a fit constant.

| T  K | $\sigma_{DC}$ (ohm$^{-1}$-cm$^{-1}$) | $\eta_1$ | $\varpi_{ph1}$ (cm$^{-1}$) | $\eta_2$ | $\varpi_{ph2}$ (cm$^{-1}$) | $\eta_3$ | $\varpi_{ph3}$ (cm$^{-1}$) |
|---|---|---|---|---|---|---|---|
| ~1624 | 87 | 7.4 | 438.3 (438.1) | 11 | 645.0 (644.4) |  |  |
| 545 | 8 |  |  | 11.2 | 707.2 (696.8) |  |  |
| 300 | 7 | 13.5 | 318.9 (326.9) | 7.6 | 735.0 (680.5) | 7.3 | 1480 |
| 105 | 7 | 7.8 | 330.8 (329.9) | 8.55 | 686.8 (682.1) | 13.7 | 1472.0 (1364.3) |
| 12 | 40 | 11.70 | 198.5 (222.8) | 15.0 | 290.1 (312.9) | 19.2 | 411.0 (421.2) |

# Table II

Parameters used in the bipolaron simulation fits for $h$-ErMnO$_3$ mid-infrared optical conductivity. $\sigma_{DC}$ in eq.(14) was used as fit fixed parameter. Corresponding experimental frequencies appear in brackets below model values.

| T (K) | 2*E$_{bipolaron}$ (cm$^{-1}$) | $\sigma_{DC}$ ($\Omega^{-1}$cm$^{-1}$) | $\omega_{(vib)}$ (cm$^{-1}$) |
|---|---|---|---|
| **545** | **3799** (**3805**) | **2.5** | **302** (**295.3**) |
| **906** | **2778** (**2509**) | **79** | **277** (**280.3**) |
| **1119** | **3775** (**3475**) | **36** | **270** (**271.4**) |
| **1298** | **3998** (**3727**) | **68** | **352.3** (**354.5**) |
| **1578** | **4784** (**4436**) | **37** | **415** (**420.1**) |
| **~1624** | **2590** (**2305**) | **86** | **412** (**438.1**) |

# FIGURE CAPTIONS

**Figure 1** (color online) (a) X-ray (CuKα) diffraction pattern for room temperature ferroelectric $h$-ErMnO$_3$ in the space group P6$_3$cm (C$^3_{6v}$), Z=6. Sharp peaks indicate very good polycrystalline quality without secondary phases. In the Rietveld plot, the red points correspond to the experimental profile and the solid line is the calculated profile, with the differences shown in the bottom blue line. ✻ signals the diffraction peaks of the trimer superstructure. The lattice structure corresponds to a view of the six-formula unit hexagonal lattice ($a$ = 6.1372(6) Å and $c$ = 11.4209(17) Å) with MnO$_5$ bipyramids plane triangular corner sharing at the nonequivalent O3 and O4 oxygen sites.[9] **(b)** Negative thermal expansion along the $c$-axis after Ref. **50**. Tc stands for the ferroelectric Curie lock-in temperature and T$_{INC}$ signals the onset of a proposed lattice incommensurate transition (see text). (c) Neutron powder diffraction profiles at ambient temperature after the Rietveld refinement: red crosses are the experimental points, the solid line is the calculated profile with the differences shown in the bottom blue line. The inset shows a view along [001] of the crystal structure, highlighting the corner sharing arrangement of MnO$_5$ units in the $a$-$b$ plane.

**Figure 2** (color online) (a) Phonon near normal reflectivity of $h$-ErMnO$_3$ between 12 K and 300 K (arrow points to the temperature evolution of the anomalous band discussed in the text). Inset: Temperature dependent far infrared reflectivity vertically offset; (b) far-infrared band tail assigned to heterogeneity from wall and domain formation (boson peak); (c) band reflectivity showing anomalous plateau step-like shoulder (see text) and phonon split at T$_N$~79 K due to spin-phonon interactions[17] and unit cell tripling.

**Figure 3** (color online) (a, b) THz transmission ratio applied field incremental of M$_1$ and M$_2$ hardening magnetic excitations in $h$-ErMnO$_3$ at 5K.[18] These are out of plane Mn moment contributions in a complex 3d and 4f interplay with Er$^{3+}$ moments at the two nonequivalent Wyckoff sites inferred in the magnetic phase diagram.[13,14] Traces have been vertically off set for better viewing. (c, d) Same data plotted as absolute ratios relative to the zero-field cooled signal. Thicker traces in (d) bring up the field dependent evolution of a net two site magnetic excitation.

**Figure 4** (color online). (a) Far infrared applied field induced absorption of $h$-ErMnO$_3$. Crystal field transitions on the 0.7 T isoline in the ferrimagnetic (2.2 K, 3.0 K) and antiferromagnetic (4.0 K, 5.0 K) phases. Traces have been vertically off set for better viewing. Notation after Ref. **18**. (b) 5 K absorption ratio of the two-site strongest Er$^{3+}$ crystal field transitions showing the doubling in energy in dotted lines either in the antiferromagnetic (1T) or ferromagnetic phase (4T, 7T).[14] (c)Temperature dependent weaker bands assigned to magnetic dressed phonons. Full line: Temperature dependent near normal reflectivity of phonon-like features (traces have been vertically off set for better viewing) ; squares: preliminary Raman scattering data at 300 K ($\lambda_{exc}$= 457 nm).

**Figure 5** (color online) (a) Phonon near normal reflectivity of $h$-ErMnO$_3$ between 298 K and 896 K. (arrow points to the temperature evolution of the anomalous band profile). Inset: Overall temperature dependent far infrared reflectivity vertically offset with arrows pointing to the softening of the anomalous band.

**Figure 6** color online) (a) Near Normal 1-Emissivity of $h$-ErMnO$_3$ between 462 K and 1502 K. Note that although at 462 K (dotted line) the sample is not hot enough to yield a less 1-E noisy spectrum it is in full agreement with phonons in Fig. 5. The spectra have been vertically offset for better viewing. (b) Overall phonon behavior from ambient to 1468 K overlapping near normal reflectivity and 1-emissivity absolute values in the common temperature range. Squares: reflectivity, full lines:1-Emissivity.

**Figure 7** (color online) (a) $h$-ErMnO$_3$ Raman spectra cross section dependence using exciting laser lines $\lambda_{exc}$= 457 nm, 514 nm, and 633 nm at 300 K. (b) Comparison of Raman spectra obtained using $\lambda_{exc}$= 355 nm and $\lambda_{exc}$=514 nm laser lines. The same sample far infrared reflectivity is outlined in the background in lighter gray trace. Inset (c) at the right: Detail showing charge carriers caused redshift (dashed vertical lines) of the Raman active LO mode using $\lambda_{exc}$= 355 nm assigned to a Fröhlich resonance. FIR reflectivity counterpart is outlined in lighter gray trace.

**Figure 8**(color online) $\sigma_1$ real part optical conductivity at 545 K, 300 K, 105 K, and 12 K in the ferroelectric phase (full line: experimental, open circle: full fit; square, upper, and lower triangles

are the individual quasi-Gaussians reproducing the experimental optical conductivities with the parameters shown in table I; the 545 K open circle envelop is the fit by bipolaron (solid starts, eq.(14)) and small polaron (eq. (10), solid square) contributions while only small polaron quasi-Gaussians, eq(10), are needed at 300 K, 105 K, and 12 K. Insets: Full spectral range of the measured dielectric function imaginary part $\varepsilon_2$.

**Figure 9** (color online) $\sigma_1$ real part optical conductivity at 906 K, 1119 K, and 1293 K in the intermediate phase (full line: experimental, open circles: fits by eq(14)). Insets: Full spectral range imaginary part $\varepsilon_2$ of the measured dielectric function. The relative increase in opacity at 1578 K, in the paraelectric phase, is in accordance with the appearance of vortex-antivortex domain structure reported at about these temperatures.[26] The real part of the optical conductivity at ~1624 K is shown in the lowest panel (full line: experimental, circle and open square : fits). The plain superposition of the two formulations, bipolaron (full line: experimental, open circles: fit by eq. (14)) and small polaron, (solid triangles: fit by eq(10)) suggests coexistence (see text). Inset: Full spectral range imaginary part $\varepsilon_2$ of the measured dielectric function.

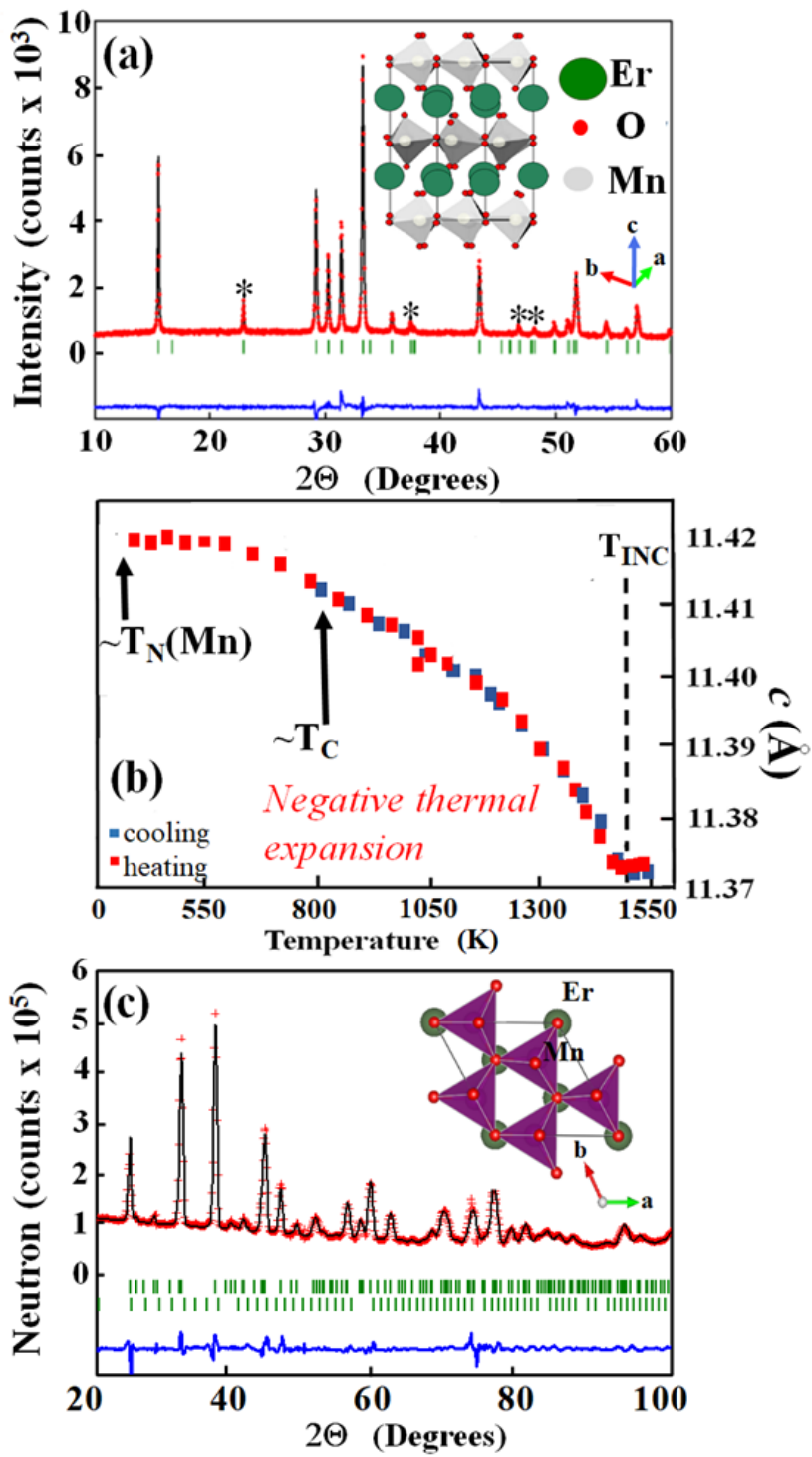

Fig. 1
Massa et al

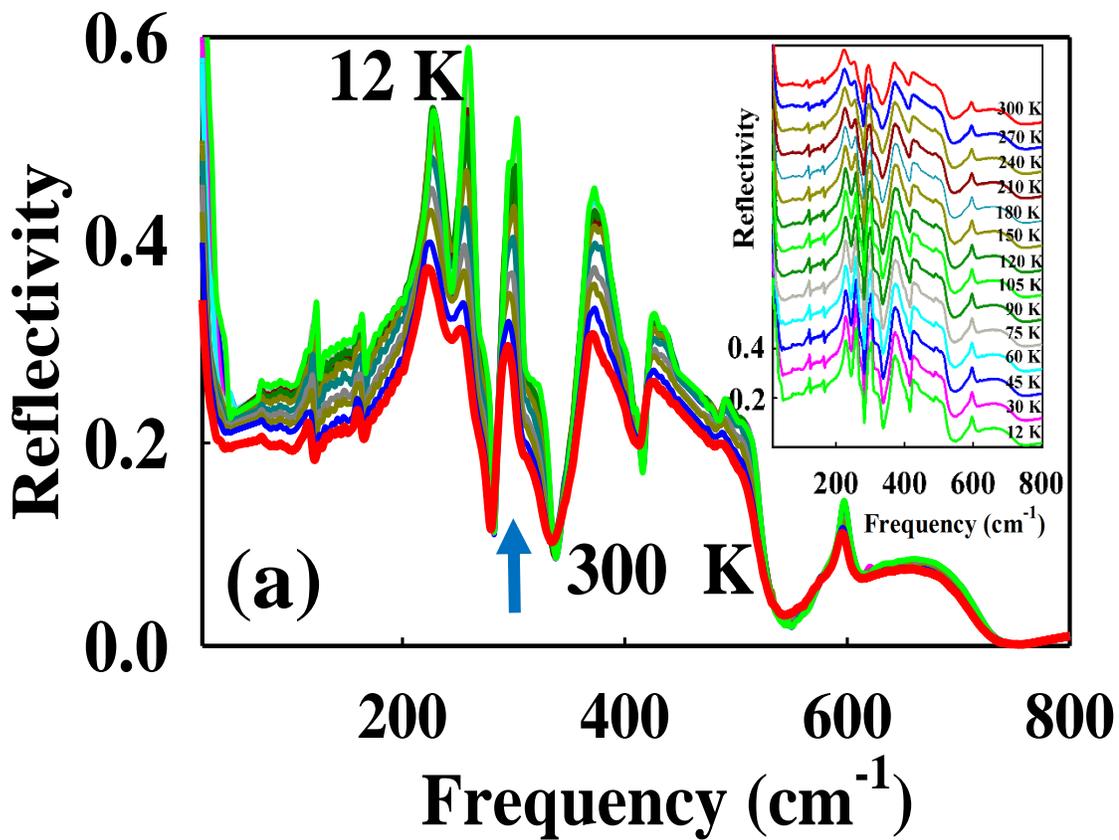
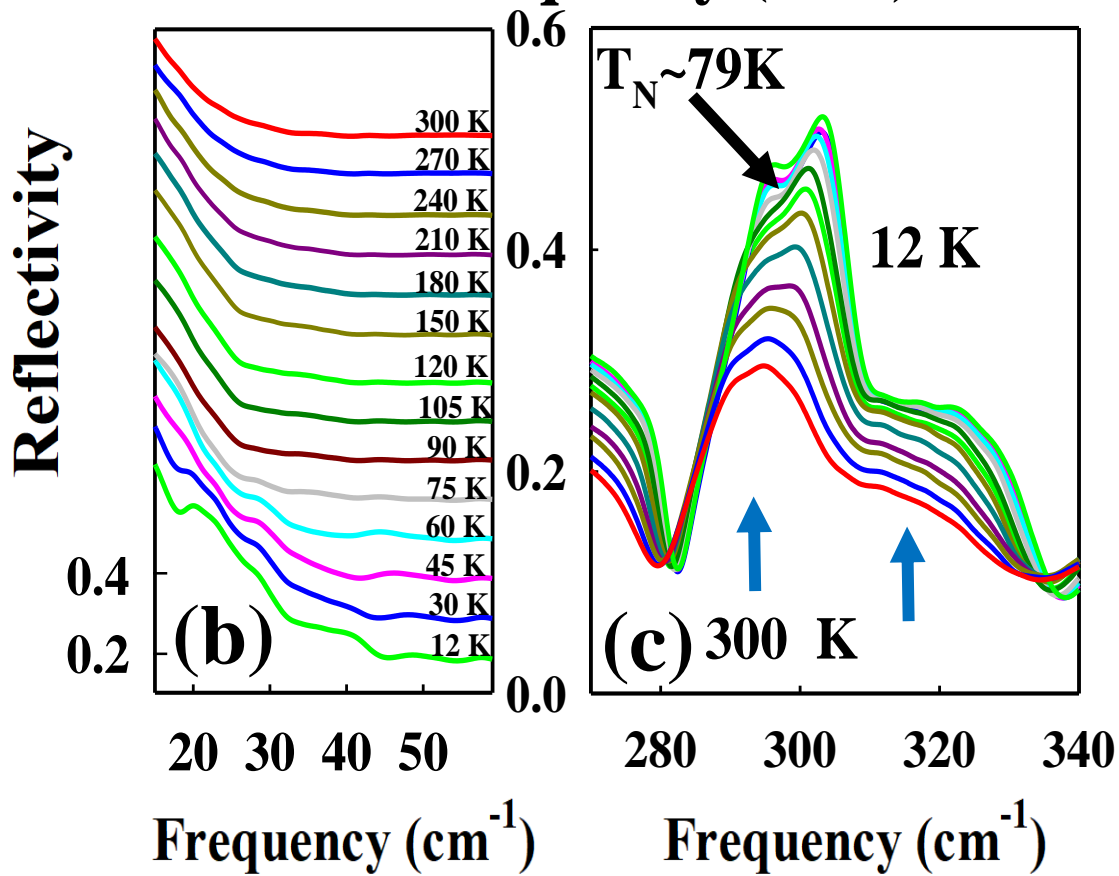

Fig. 2
Massa et al

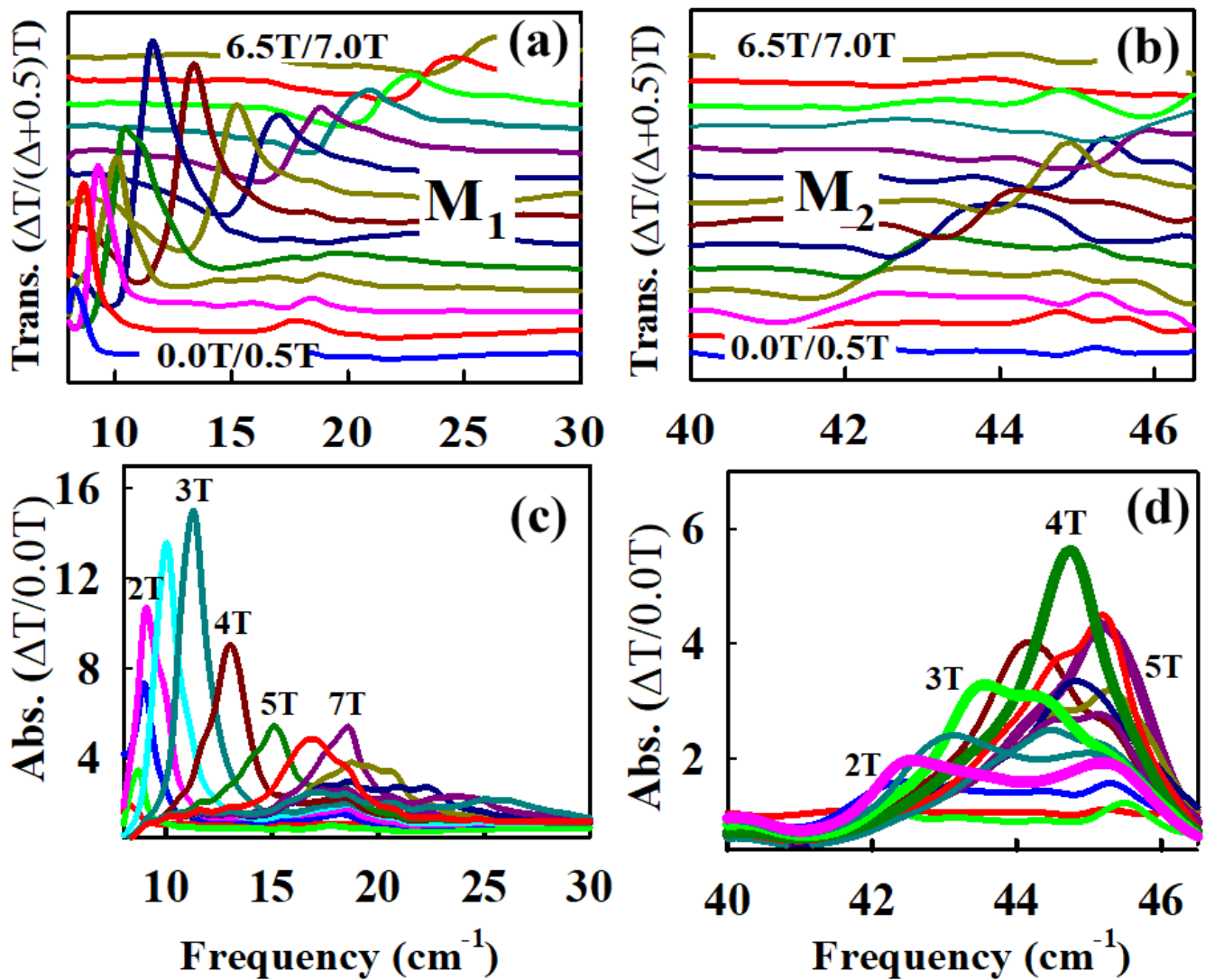

Fig. 3
Massa et al
39

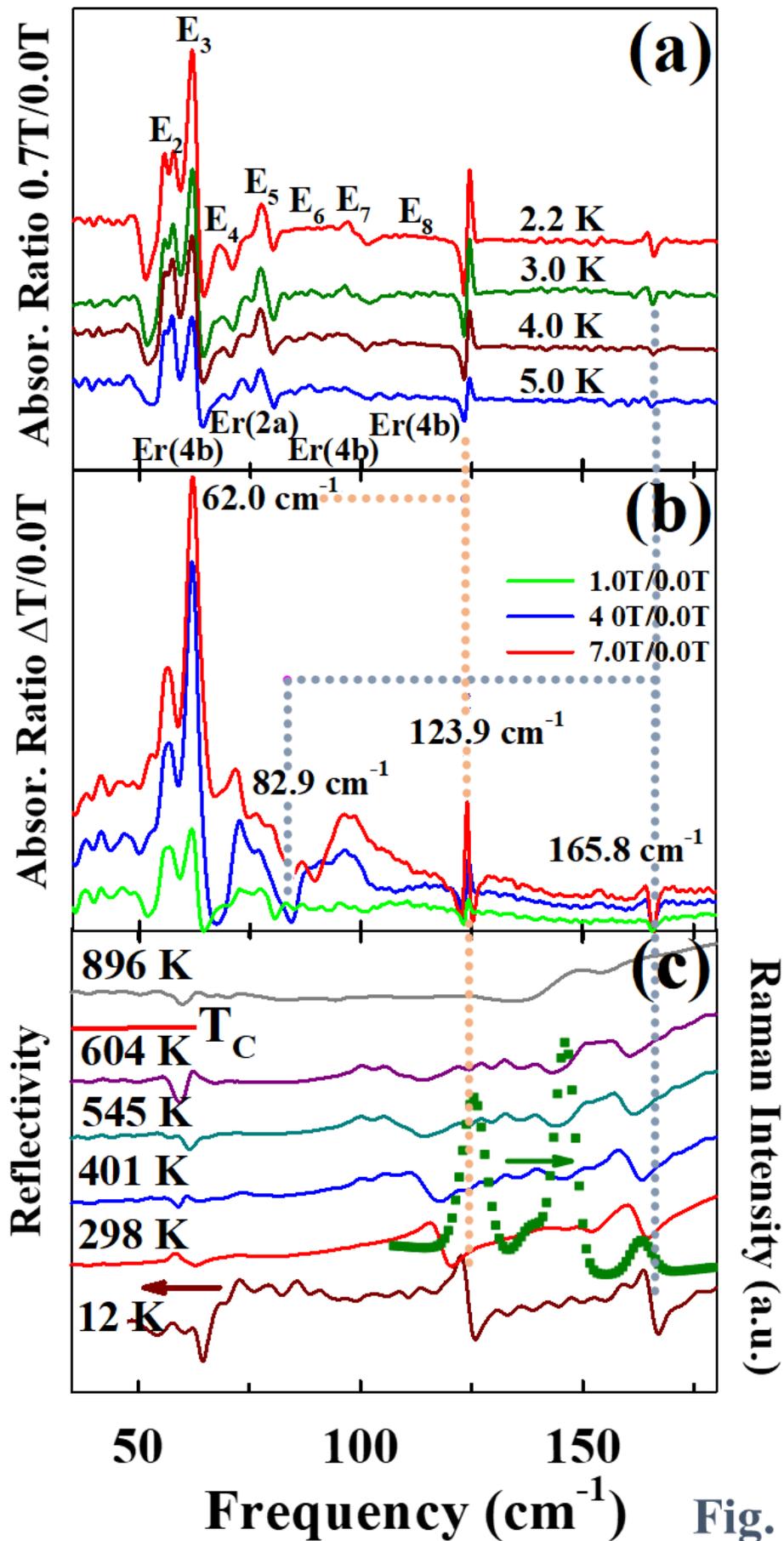

Fig. 4 Massa et al

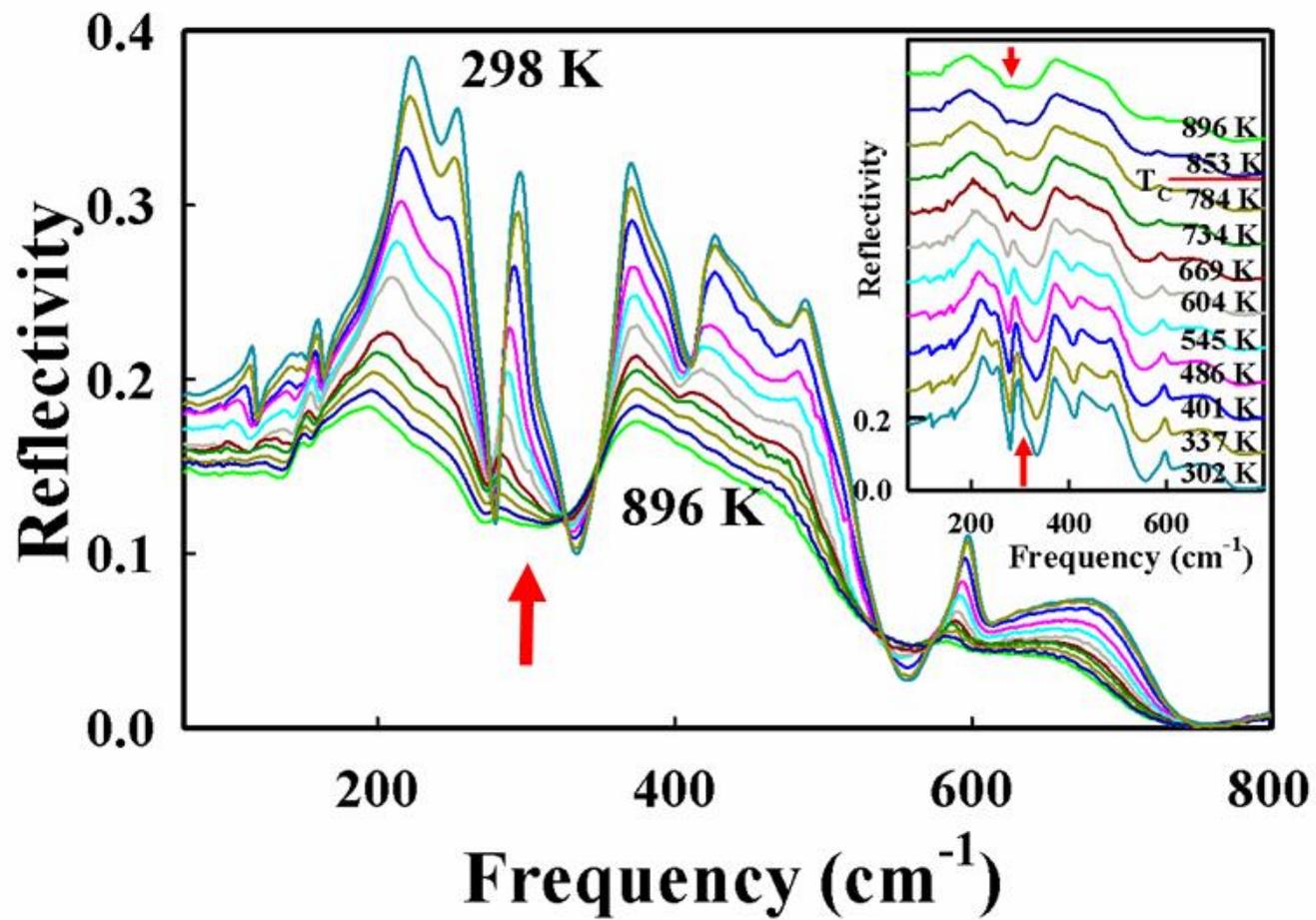

Fig. 5
Massa et al

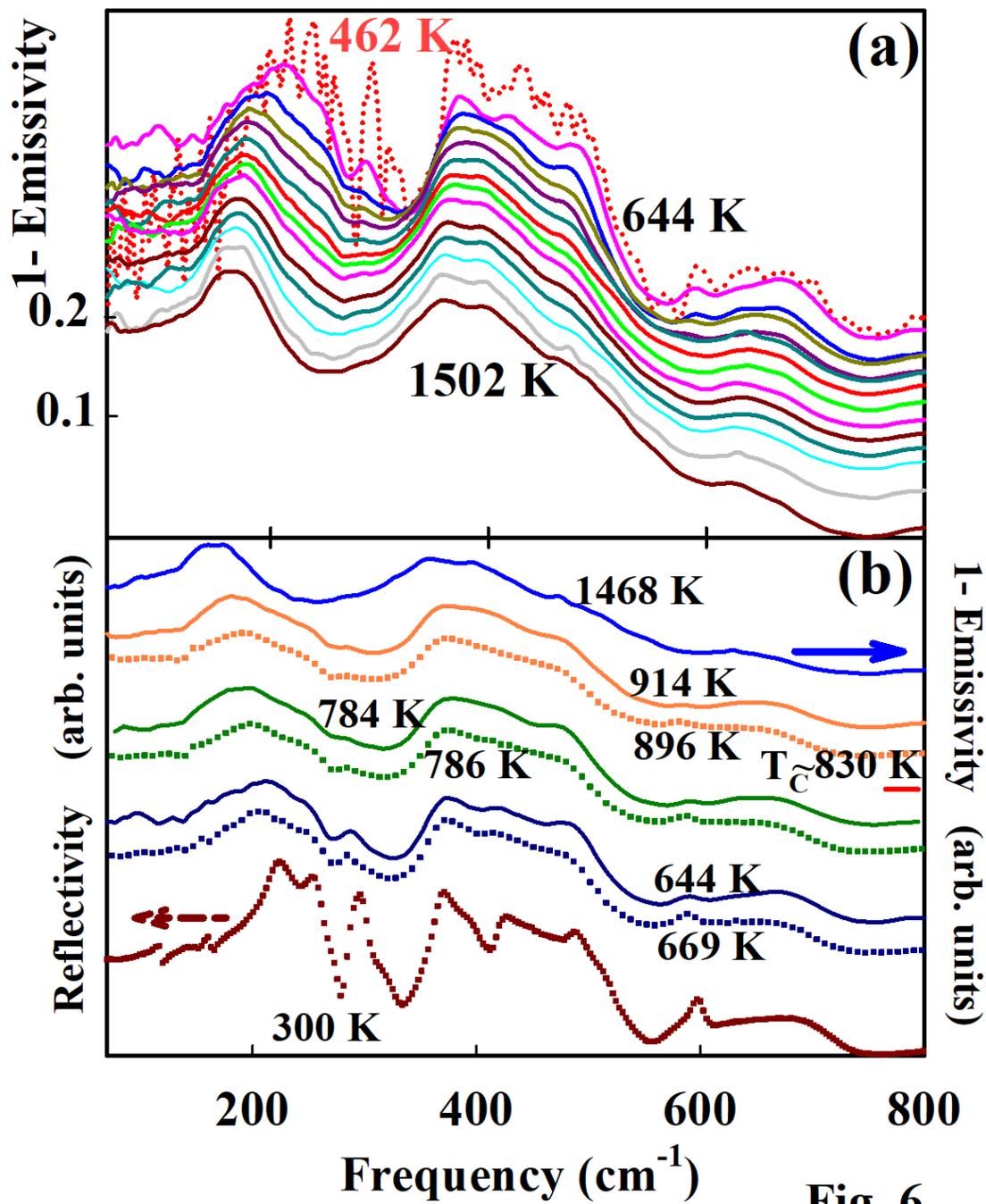

Fig. 6
Massa et al

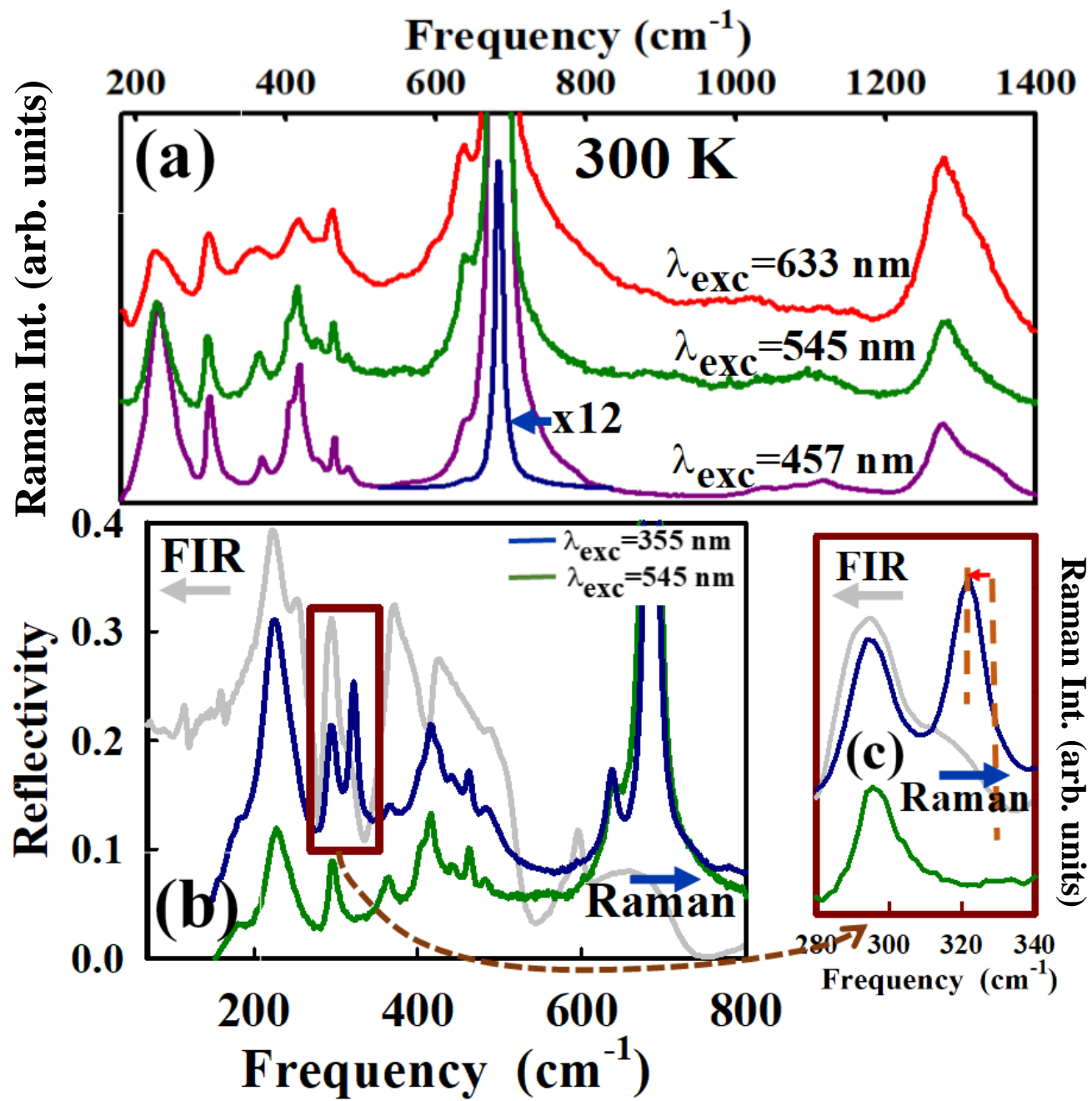

Fig. 7 Massa et al

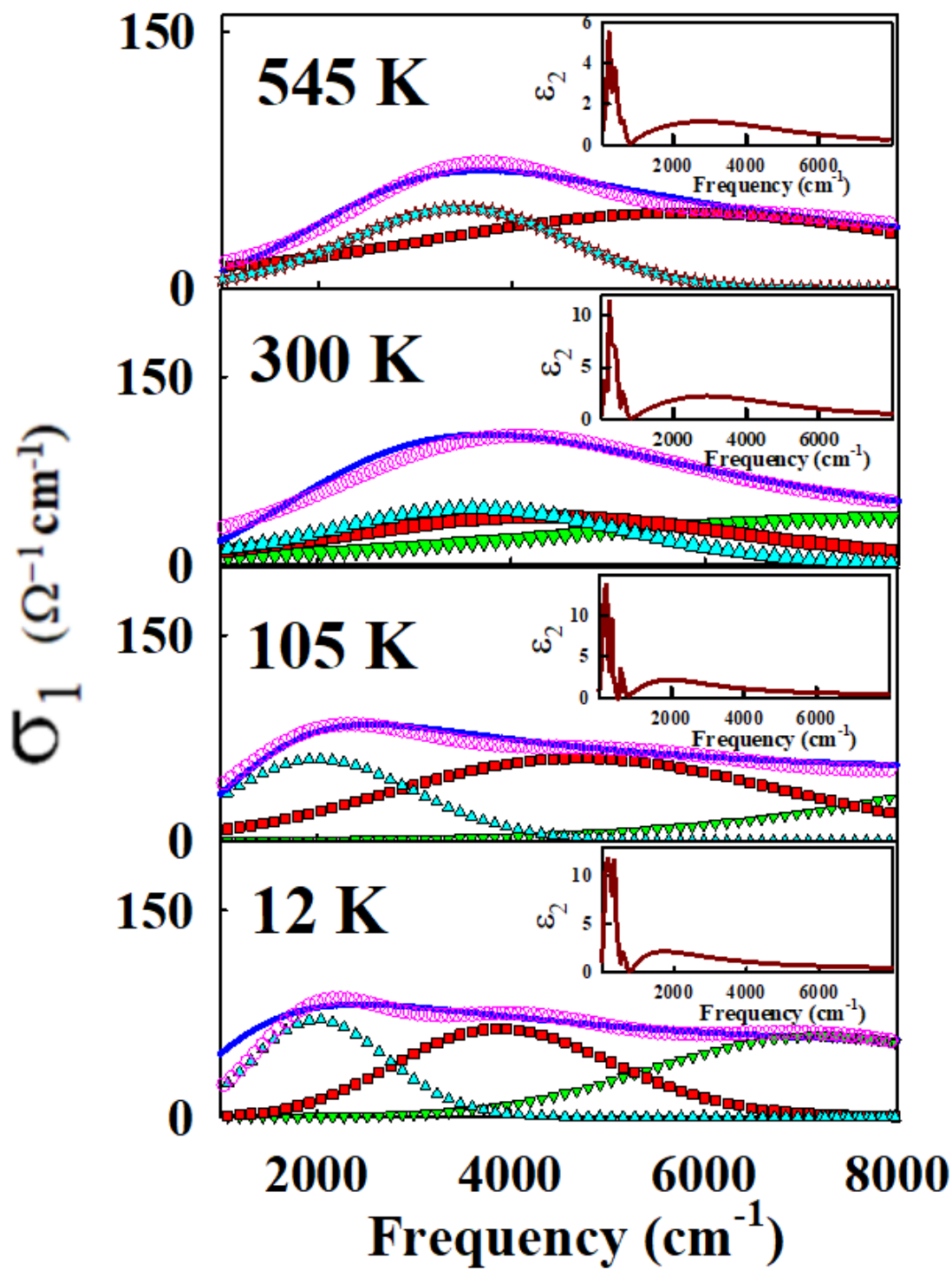

Fig. 8
Massa et al

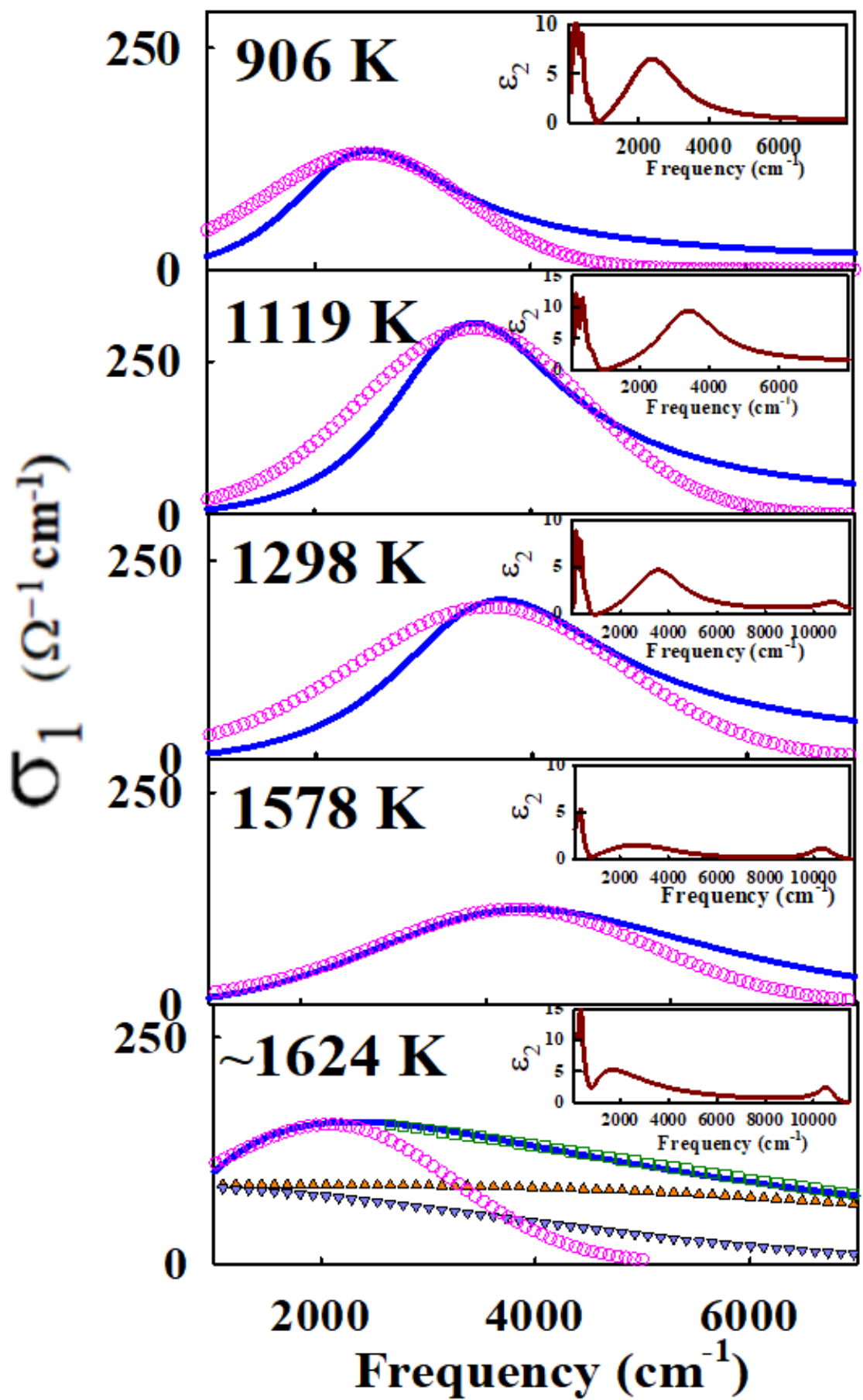

Fig. 9
Massa et al

# SUPPLEMENTAL MATERIAL

# $h$-ErMnO$_3$ absorbance, reflectivity and, emissivity in the THz to mid-infrared from 2 K to 1700 K: carrier screening, Fröhlich resonance, small polarons, and bipolarons

Néstor E. Massa,*,[1] Leire del Campo,[2] Karsten Holldack,[3] Aurélien Canizarès,[2] Vinh Ta Phuoc,[4] Paula Kayser,[5] and José Antonio Alonso[6]

[1] Centro CEQUINOR, Consejo Nacional de Investigaciones Científicas y Técnicas, Universidad Nacional de La Plata, Bv. 120 1465, B1904 La Plata, Argentina.

[2] Centre National de la Recherche Scientifique, CEMHTI UPR3079, Université Orléans, F-45071 Orléans, France

[3] Helmholtz-Zentrum für Materialien und Energie GmbH,) Albert Einstein Str. 15, D-12489 Berlin, Germany.

[4] Groupement de Recherche Matériaux Microélectronique Acoustique Nanotechnologies- Université François Rabelais Tours, Faculté des Sciences & Techniques, F- 37200 Tours, France.

[5] Centre for Science at Extreme Conditions and School of Chemistry, University of Edinburgh, Kings Buildings, Mayfield Road, EH9 3FD Edinburgh, United Kingdom.

[6] Instituto de Ciencia de Materiales de Madrid, CSIC, Cantoblanco, E-28049 Madrid, Spain.

•e-mail: neemmassa@gmail.com

# Sample preparation and structural characterization

$h$-ErMnO$_3$ was prepared as polycrystalline powder by a liquid-mix technique. Stoichiometric amounts of analytical-grade Er$_2$O$_3$ and MnCO$_3$ were dissolved in citric acid by adding several droplets of concentrated HNO$_3$ to favor the solution of Er$_2$O$_3$; the citrate + nitrate solution was slowly evaporated, leading to an organic resin that was first dried at 120 °C, and then, decomposed by heating at temperatures up to 800 °C in air. The precursor powder was finally heated at 1100 °C in air for 12 h, thus yielding a well-crystallized single- phase powder that were used for sinter pellet making. Fig. 1(a) shows the laboratory XRD pattern.

We want to stress that our measurements have been performed in samples carefully characterized by high-resolution neutron diffraction patterns (NDP), which assess their quality (well crystallized and with perfectly modeling crystal structures). With respect to precedent reports, all performed from x-ray diffraction data, the present characterization offers a more reliable accuracy regarding the oxygen positions, better determined by neutrons. NDP data were collected at ILL-Grenoble, in the high-flux D20 line with take-off angle of 42° and a copper Cu (200) monochromator with optimized fixed vertical focusing giving a wavelength of 1.30 Å. At this wavelength, the monochromatic beam has its highest flux of about 9.8 x 10$^7$ n cm$^{-2}$ s$^{-1}$. The detector was a large microstrip PSD multidetector with 1536 detection cells covering 153.6°.

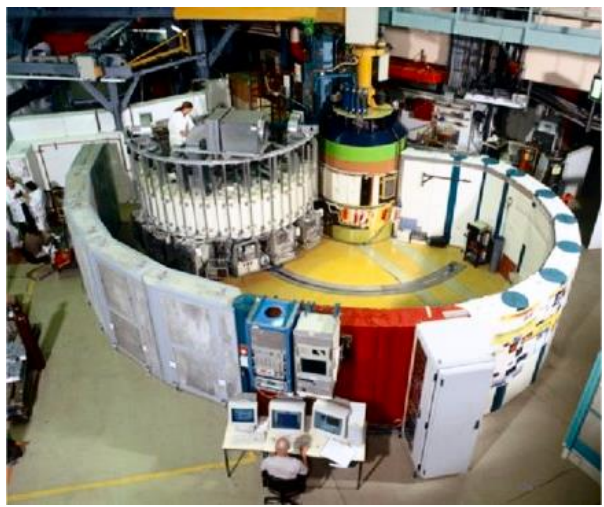

**Fig. S1** D20 –High Intensity two-axis diffractometer located at the Institut Laue–Langevin (ILL-Grenoble) reactor hall, thermal beam H11.

The initial model for the refinement of the $h$-ErMnO$_3$ crystal structure was taken from Ref. 1. It is defined in the hexagonal space group $P6_3cm$ ($C^3_{6v}$); the refined lattice parameters are $\underline{a}$ = 6.1372(6) Å and $\underline{c}$ = 11.4209(17) Å. Mn$^{3+}$ ions are at Wyckoff sites **6c**, the oxygens lie at 3 sites **6c, 2a** and **4b**, while the two Er$^{3+}$ are at the two non-equivalent crystallographic sites, **4b**, and **2a**.[2]

Table 1 lists the structural parameters refined from NDP data at RT. As it is shown in Fig. 1a each Mn atom is coordinated by five oxygen atoms in a bipyramidal configuration. One O3 atom and two O4 atoms are in the equatorial plane of the bipyramid, whereas the O1 and O2 atoms are at the apexes. Er occupies two crystallographic positions, bonded each to seven oxygen atoms. Both $ErO_7$ polyhedra may be labeled as monocapped octahedra. The capping oxygen is O3 for Er1 and O4 for Er2. Along the **c** axis, the structure consists of layers of corner-sharing $MnO_5$ bipyramids separated by layers of edge-sharing $ErO_7$ polyhedra.[2] Fig. 1(c) displays the quality of the fit from NPD data, and the inset contains a view along the **c** axis of one layer of $MnO_5$ units sharing corners forming a characteristic triangular arrangement.

## Table SI

Fractional atomic coordinates and isotropic displacement parameters ($Å^2$) for Hexagonal $ErMnO_3$, $P6_3cm$, with a = 6.1372(6) Å and c = 11.4209(17) Å, V= 372.53 (8) $Å^3$.

|     | x          | y       | z          | $U_{iso}$    |
|-----|------------|---------|------------|--------------|
| Er1 | 0.00000    | 0.00000 | 0.2707 (14)| 0.00532(2)   |
| Er2 | 0.33333    | 0.66666 | 0.2289 (12)| 0.00532(2)   |
| Mn  | 0.32130    | 0.00000 | 0.00000    | 0.0067 (14)  |
| O1  | 0.3048 (14)| 0.00000 | 0.162 (2)  | 0.0055 (6)   |
| O2  | 0.6403 (14)| 0.00000 | 0.338 (2)  | 0.0055 (6)   |
| O3  | 0.00000    | 0.00000 | 0.472 (2)  | 0.0055 (6)   |
| O4  | 0.33333    | 0.66666 | 0.0130 (19)| 0.0055 (6)   |

# Experimental

Low temperature-low frequency absorbance measurements from 3 cm$^{-1}$ to 50 cm$^{-1}$ with 0.5 cm$^{-1}$ resolution have been performed in the THz beamline of the BESSY II storage ring at the Helmholtz-Zentrum Berlin (HZB) in the low-alpha multi bunch hybrid mode. In the synchrotron low-alpha mode electrons are compressed within shorter bunches of only ~2 ps duration allowing far-infrared wave trains up to mW average power to overlap coherently in the THz range below 50 cm$^{-1}$. We used a superconducting magnet (Oxford Spectromag 4000, here up 7.5 T) interfaced with the interferometer for the measurements under magnetic fields.[3] Measurements in the 30 cm$^{-1}$ to 300 cm$^{-1}$ range were also taken using the internal source of a Bruker IFS125 HR. The temperature was measured with a calibrated Cernox Sensor from LakeShore Cryotronics mounted to the copper block that holds the sample in the Variable Temperature Insert (VTI) of the Spectromag 4000 Magnet. To avoid masking fundamental interactions by signal saturation we also run far infrared spectra of $h$-ErMnO$_3$ thin powder coatings. These lasts were made of powder stick on a scotch tape after pressuring it on the surface of a rough polishing cloth loaded after rubbing on it an *h*-ErMnO$_3$ pellet. The coatings that weighted about 40 µgr resulted of ideal thickness for low temperature transmission measurements in the range from 30 cm$^{-1}$ to 180 cm$^{-1}$.

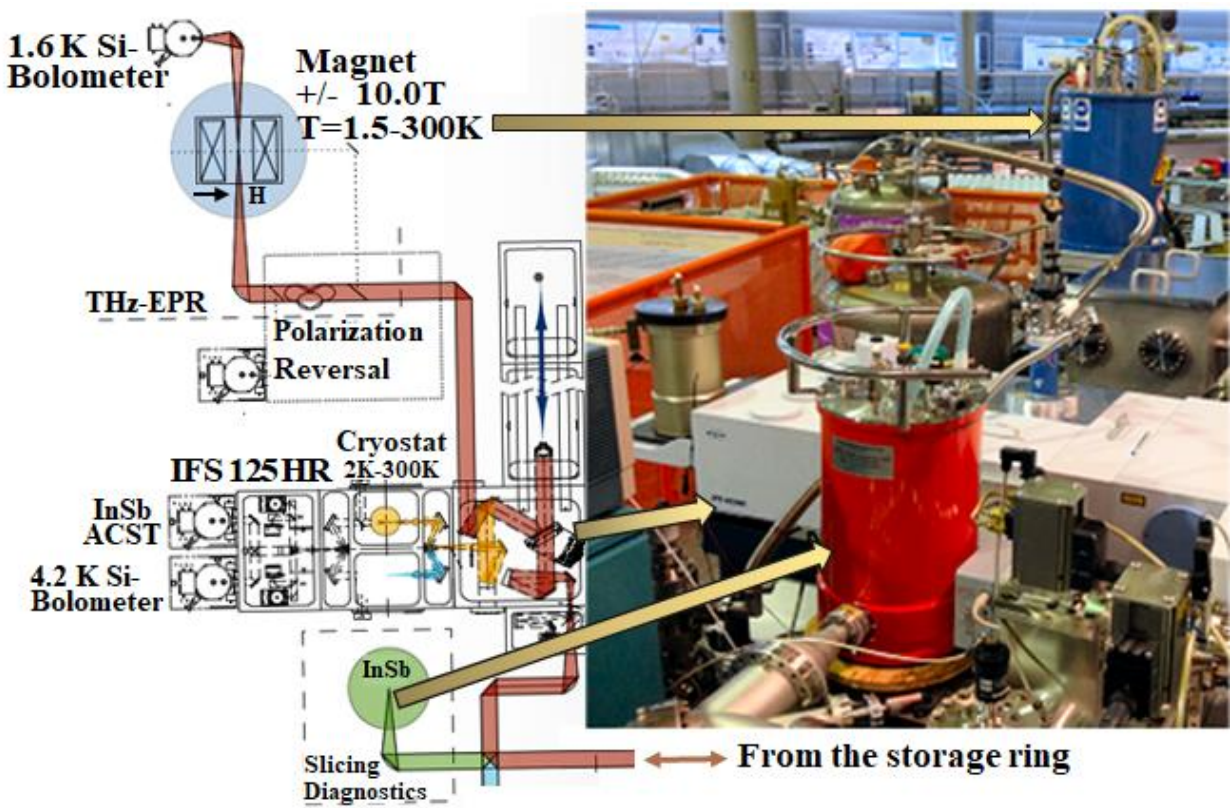

**Fig. S2.** Beam optical path and layout of the THz beamline at BESSY II (Ref. 3).

Our preliminary Raman data were taken at 1 cm$^{-1}$ resolution with a Qontor l spectrometer equipped with an Leica 100x microscope objective (NA 0.85) with a 1800 gr/mm
and beam optical path grating with the 457 nm (Coherent Sapphire single frequency) and 545 nm (Cobolt Fandango™) exciting lines. The same set up was used with λ$_{exc}$ =355nm but replacing the 1800 gr/mm by the 3600 gr/mm grating and a Thorlabs LMU-40X-NUV (NA 0.5) focusing objective. The triple subtractive configuration (1800 gr/mm gratings) of the Horiba Jobin Yvon T64000 spectrometer allowed Raman acquisition in the low-frequency range up to ~100 cm$^{-1}$. In every case laser power on the sample was always less than 1 mW.[4]

Infrared spectra were taken on heating using two experimental facilities, one corresponds to reflection and a second one for emission. From 12 K to room temperature and from 300 K up to about ~850K we have measured reflectivity at 1 cm$^{-1}$ resolution with two Fourier transform infrared spectrometers, a Bruker 66 v/s, and a Bruker 113V, respectively, with conventional near normal incidence geometries. Low temperature runs were made with the sample mounted in the cold finger of a Displex closed cycle He refrigerator. For high temperature reflectivity a heating plate adapted to the near normal reflectivity attachment was used in the Bruker 113v vacuum chamber. A liquid He cooled bolometer and a deuterated triglycine sulfate pyroelectric bolometer (DTGS) were employed to completely cover the spectral range of interest. For the low temperature set an evaporated in situ gold film was used as 100% reference reflectivity while a plain gold mirror was used for reflectivity between 300 K and 800 K. In this temperature range, the spurious infrared signal introduced by the hot sample thermal radiation was corrected to obtain the absolute reflectivity values.

Near normal emissivity at 2 cm$^{-1}$ resolution was measured with two Fourier transform infrared spectrometers, Bruker Vertex 80v and Bruker Vertex 70, coupled to a rotating table placed inside a dry air box allowing to simultaneously measure the spectral emittance in two dissimilar spectral ranges from 40 cm$^{-1}$ to 13000 cm$^{-1}$.[5, 6] In addition to the infrared region covered by a DTGS detector, the spectral regions of interest were the far infrared from 40 cm$^{-1}$ to 900 cm$^{-1}$ using a He cooled bolometer, and another from 7000 cm$^{-1}$ to 13000 cm$^{-1}$ with a high gain InGaAs photodiode.

The sample, which was heated with a 500 W pulse Coherent $CO_2$ laser, was positioned on the rotating table at the focal point of both spectrometers in a position equivalent to that of the internal radiation sources inside the spectrometers. In this measuring configuration, the sample, placed outside the spectrometer, is the infrared radiation source, and conversely, the sample chamber inside the spectrometers is empty. To assure reliability and reproducibility, since the radiation at highest heating powers might induce extrinsic defects, the shown measurements in the ~500 K to ~1700 K range are result of about 10 independent runs using a fresh pellet in every case. In oxides heating samples at the highest temperatures in dry air may induce additional conductivity due to double exchange by $Mn^{3+}$ and $Mn^{4+}$ ions. It is known that annealing oxides in air freer electrons may appear consequence of the oxidation process $Mn^{3+} \rightarrow Mn^{4+} + 1e-$ , which is equivalent to divalent $A^{2+}$ substitution, creating a higher-mobility-high-temperature small polaron environment.[8]

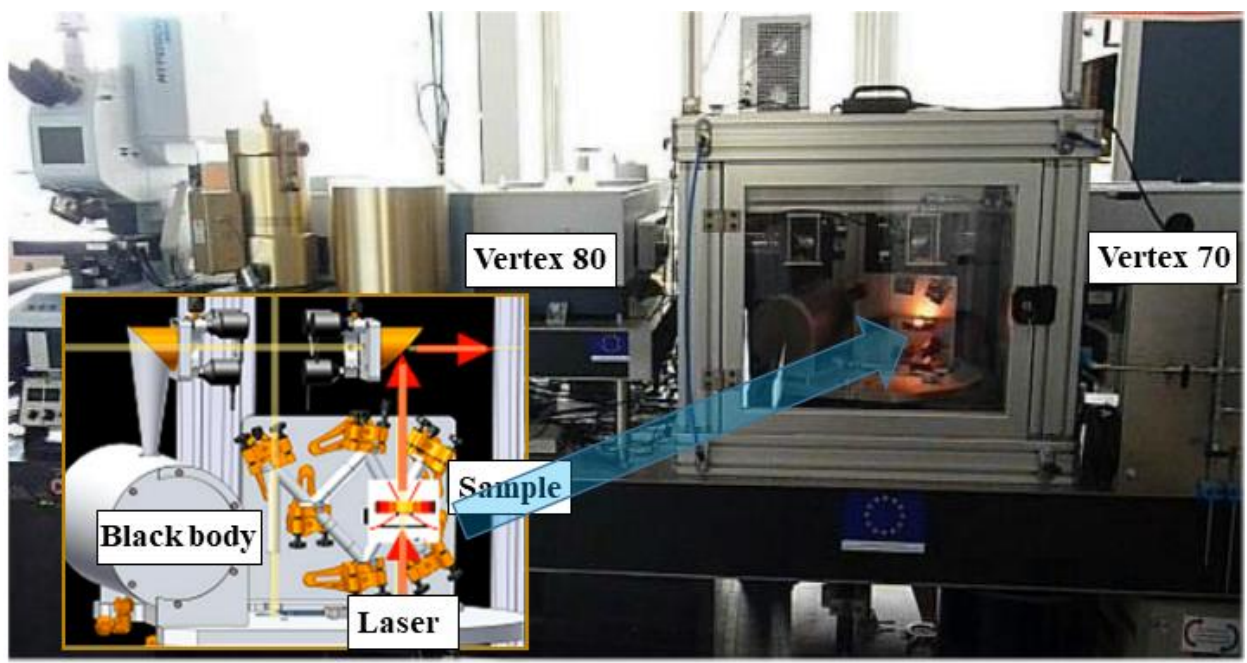

**Fig. S3** Infrared emission set up showing the infrared spectrometers Vertex 70 and 80v allowing emittance measurements in the spectral range from far infrared to visible (30-17000 cm$^{-1}$) and up to temperatures of 2500 K. Inset: Heating plate using the $CO_2$ laser. The acquisition of the sample flux is limited to a small area of the sample (2 mm of diameter) to discard heat radial gradients.[7]

# METHODS AND DATA ANALYSIS

We reiterate in the following lines basic concepts and tools used in our data analysis already discussed elsewhere.[5, 6, 9]

Normal spectral emissivity of a sample, $\mathbf{E}(\omega,T)$, is given by the ratio of its luminescence ($L_S$) relative to the black body's luminescence ($L_{BB}$) at the same temperature $T$ and geometrical conditions, thus,

$$\mathbf{E}(\omega,T) = \frac{L_S(\omega,T)}{L_{BB}(\omega,T)} \qquad (S1)$$

In practice, the evaluation of this quantity needs the use of a more complex expression because the measured fluxes are polluted by parasitic radiation. This is because part of the spectrometer and detectors are at 300 K. To eliminate this environmental contribution the sample emissivity is retrieved from three measured interferograms, i.e.

$$\mathbf{E}(\omega,T) = \frac{FT(I_S - I_{RT})}{FT(I_{BB} - I_{RT})} \times \frac{\mathscr{P}(T_{BB}) - \mathscr{P}(T_{RT})}{\mathscr{P}(T_S) - \mathscr{P}(T_{RT})} \mathbf{E}_{BB} \qquad (S2)$$

where $FT$ stands for Fourier Transform, and $I$ for measured interferogram i.e., sample, $I_S$; black body, $I_{BB}$; and, environment, $I_{RT}$. $\mathscr{P}$ is the Planck's function taken at different temperatures T; i.e., sample, $T_S$; blackbody, $T_{BB}$; and surroundings, $T_{RT}$. $\mathbf{E}_{BB}$ is a correction that corresponds to the normal spectral emissivity of the black body reference (a $LaCrO_3$ Pyrox PY 8 commercial oven) and takes into account its non-ideality.[5]

One of the advantages of emissivity is that allows contact free measurement of the temperature of a measured insulator using the Christiansen frequency; i.e., the frequency where the refraction index is equal to one, and the extinction coefficient is negligible, just after the highest longitudinal optical phonon frequency. The temperature is calculated with emissivity $E(\omega, T)$, eq. (S2), set equal to one at the Christiansen frequency. Since this varies its position with temperature after identifying the frequency in which $E(\omega_{Christ}, T_{sample}) = 1$ the only remaining variable left is $T_{sample}$, the sample temperature that univocally corresponds to it.

In addition, we imposed as extra condition that sample emission cannot be higher than one in spectra of poor conducting samples where there is a weak but significant Drude term (Fig. S10). This is because the Christiansen point becomes a less reliable temperature reference since the extra presumably hopping carriers may conceal or distort the actual frequency minimum.

Then, in a regular run, one interferometer always measures the infrared DTGS region, where the Christiansen point lies, while the other covers any other spectral region of interest, both, focused at the same sample spot. This is the reason of having two interferometers measuring simultaneously.

After acquiring the optical data, we placed our spectra in a more familiar near normal reflectivity framework using the second Kirchhoff law, that is,

$$R = 1 - E \qquad (S3)$$

where $R$ is the sample reflectivity. It assumes that in the spectral range of interest any possible transmission is negligible being emissivity just the complement of reflectivity.

This allows computing reflectivity and emissivity oscillator frequencies using a standard multioscillator dielectric simulation[10, 11] with the dielectric function, $\varepsilon(\omega)$, given by

$$\varepsilon(\omega) = \varepsilon_1(\omega) - i\varepsilon_2(\omega) = \varepsilon_\infty \prod_j \frac{(\omega_{jLO}^2 - \omega^2 + i\gamma_{jLO}\omega)}{(\omega_{jTO}^2 - \omega^2 + i\gamma_{jTO}\omega)} \quad (S4)$$

$\varepsilon_\infty$ is the high frequency dielectric constant taking into account electronic contributions; $\omega_{jTO}$ and $\omega_{jLO}$, are the transverse and longitudinal optical mode frequencies and $\gamma_{jTO}$ and $\gamma_{jLO}$ their respective damping. We also added to the dielectric simulation, when needed, a double damping extended Drude term (correlated plasma contribution),

$$-\frac{\left(\omega_{pl}^2 + i \cdot (\gamma_{pl} - \gamma_0) \cdot \omega\right)}{(\omega - i\gamma_0) \cdot \omega} \quad \text{(S5)}$$

where $\omega_{pl} = \sqrt{\frac{n\,e^2}{\varepsilon_0 \varepsilon_\infty m^*}}$ is the plasma frequency (*n*, the number of carriers, *e*, the carrier charge; *m\**, the effective mass). $\gamma_{pl}$ its damping, and $\gamma_0$ (the inverse of relaxation time at zero frequency) understood as a phenomenological damping introduced to reflect lattice drag effects. When these two dampings are set equal, one retrieves the classical Drude formula.[12]

The real *(ε₁(ω))* and imaginary *(ε₂(ω))* part of the dielectric function (complex permittivity, *ε\*(ω)*) is then estimated from fitting the data using the reflectivity R given by[13]

$$R(\omega) = \left| \frac{\sqrt{\varepsilon^*(\omega)} - 1}{\sqrt{\varepsilon^*(\omega)} + 1} \right|^2 \quad \text{(S6)}$$

We also calculated the real part of the temperature dependent optical conductivity, $\sigma_1(\omega)$ that given by[13]

$$\sigma_1(\omega) = \frac{\omega \cdot \varepsilon_2(\omega)}{4\pi} \quad \text{(S7)}$$

constitutes our experimental measured optical conductivity.

# Temperature Dependent Reflectivity and 1-Emissivity and Multioscillator Fits

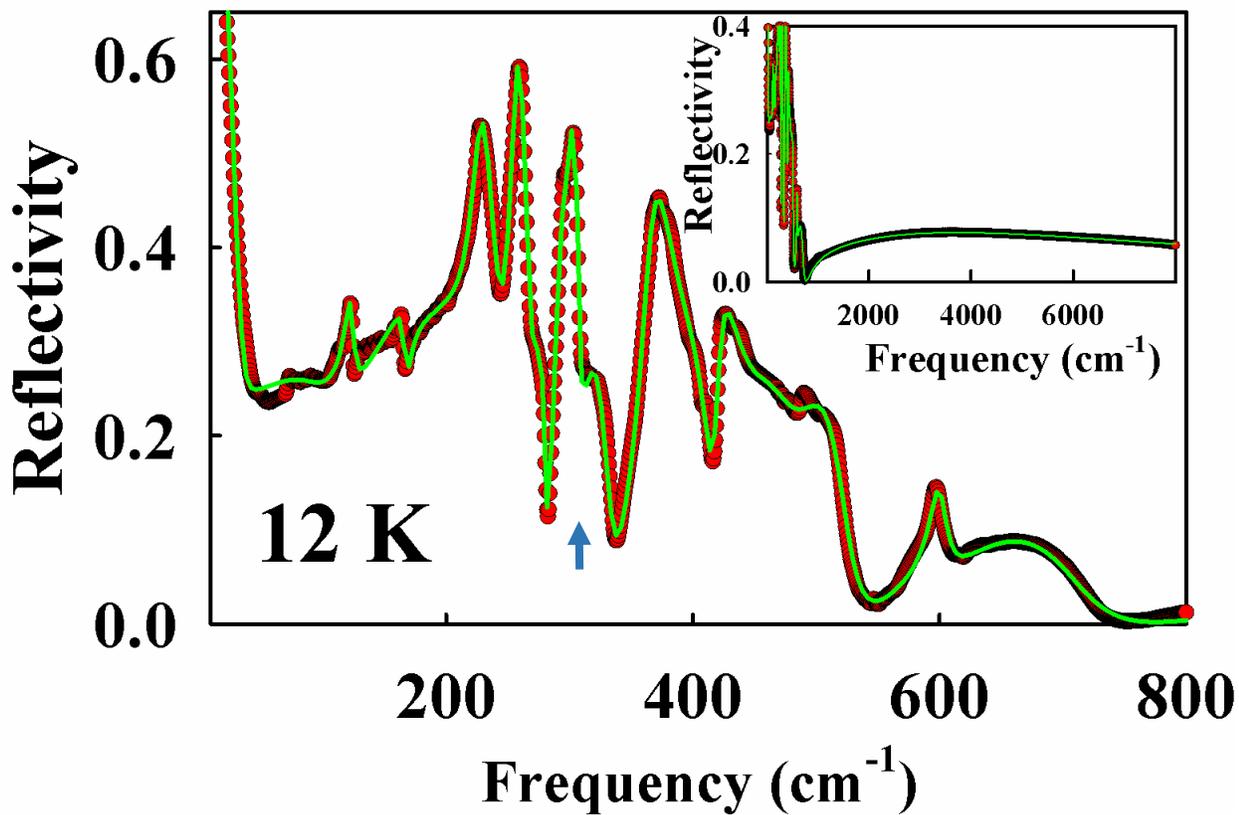

**Fig. S4** *h*-ErMnO$_3$ near normal reflectivity at 12 K, experimental: dots, full line: fit. Inset: full measured reflectivity range. Note that in the MIR part of these spectra to facilitate the fit we deleted second order phonons and weak Er$^{3+}$ crystal field substructure. Arrow points to the bands with anomalous temperature dependent profile. Fitting parameters are shown in Table SII.

# Table SII

Dielectric simulation fitting parameters for *h*-ErMnO$_3$ reflectivity at 12 K. Bottom cells show the parameters used for the lower frequency Gaussian fit shown in Fig. S4 Anomalous phonon frequencies appearing in the polaron fits are indicated in bold italics with an asterisk.

| T (K) | $\varepsilon_\infty$ | $\omega_{TO}$ (cm$^{-1}$) | $\Gamma_{TO}$ (cm$^{-1}$) | $\omega_{LO}$ (cm$^{-1}$) | $\Gamma_{LO}$ (cm$^{-1}$) |
|---|---|---|---|---|---|
| 12 | 1.55 | 91.5 | 86.3 | 96.7 | 77.2 |
| | | 121.1 | 8.9 | 122.5 | 12.2 |
| | | 164.9 | 12.8 | 166.6 | 10.3 |
| | | 186.9 | 40.2 | 193.2 | 41.2 |
| | | 228.1 | 21.5 | 239.0 | 14.9 |
| | | 254.5 | 16.3 | 266.3 | 11.9 |
| | | 278.2 | 17.8 | 280.8 | 6.3 |
| | | ***\*292.7*** | ***11.7*** | ***302.6*** | ***13.0*** |
| | | ***\*302.7*** | ***6.5*** | ***305.4*** | ***9.8*** |
| | | ***\*322.4*** | ***35.0*** | ***333.1*** | ***15.9*** |
| | | 362.8 | 21.0 | 378.1 | 49.0 |
| | | 401.3 | 110.3 | 416.1 | 18.5 |
| | | 420.3 | 11.8 | 423.4 | 62.3 |
| | | 461.6 | 81.6 | 479.6 | 56.9 |
| | | 503.1 | 62.6 | 521.4 | 24.8 |
| | | 595.3 | 19.0 | 600.6 | 17.7 |
| | | 660.8 | 211.4 | 705.8 | 66.5 |
| | | 2005.4 | 4063.8 | 5689.9 | 8048.1 |
| | | Peak position $\omega_{0G}$ | Bandwidth | | A |
| | | 7.00 | W=19.3 | | A=1334.1 |

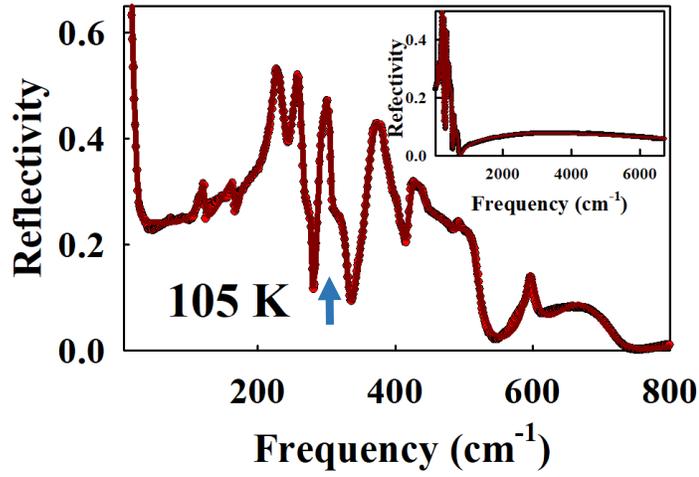

**Fig. S5** *h*-ErMnO$_3$ near normal reflectivity at 105 K, experimental: dots, full line: fit. Inset: full measured reflectivity range. Note that in the MIR part of these spectra to facilitate the fit we deleted second order phonons and weak Er$^{3+}$ crystal field substructure. Arrow points to the bands with anomalous temperature dependent profile. Fitting parameters are shown in Table SIII

## Table SIII

Dielectric simulation fitting parameters for *h*-ErMnO$_3$ reflectivity at 105 K. Bottom cells show the parameters used for the lower frequency Gaussian fit shown in Fig. S5. Anomalous phonon frequencies appearing in the polaron fits are indicated in bold italics with an asterisk.

| T (K) | $\varepsilon_\infty$ | $\omega_{TO}$ (cm$^{-1}$) | $\Gamma_{TO}$ (cm$^{-1}$) | $\omega_{LO}$ (cm$^{-1}$) | $\Gamma_{LO}$ (cm$^{-1}$) |
|---|---|---|---|---|---|
| 105 | 1.52 | 98.5 | 139.0 | 101.2 | 126.5 |
| | | 120.3 | 8.6 | 121.8 | 10.2 |
| | | 165.8 | 11.9 | 166.8 | 8.7 |
| | | 183.1 | 48.8 | 193.1 | 53.4 |
| | | 224.7 | 21.1 | 237.5 | 20.7 |
| | | 254.7 | 22.4 | 265.7 | 12.8 |
| | | 278.4 | 17.6 | 280.1 | 5.5 |
| | | ***289.1*** | ***10.5*** | ***295.6*** | ***20.8*** |
| | | ***301.7*** | ***11.7*** | ***308.4*** | ***9.7*** |
| | | ****322.6*** | ***43.6*** | ***330.1*** | ***15.5*** |
| | | 361.7 | 30.9 | 383.4 | 41.0 |
| | | 401.5 | 97.0 | 415.2 | 16.7 |
| | | 418.3 | 13.4 | 441.7 | 67.2 |
| | | 470.3 | 60.2 | 477.0 | 56.4 |
| | | 494.1 | 187.8 | 513.5 | 16.7 |
| | | 596.6 | 12.1 | 598.9 | 32.8 |
| | | *658.1 | 97.5 | 712.4 | 76.9 |
| | | 2132.5 | 3375. | 5380.8 | 7470.3 |
| | | A=967.4 | | W=16.98 | |
| | | | $\omega_{0G}$=2.62 | | |

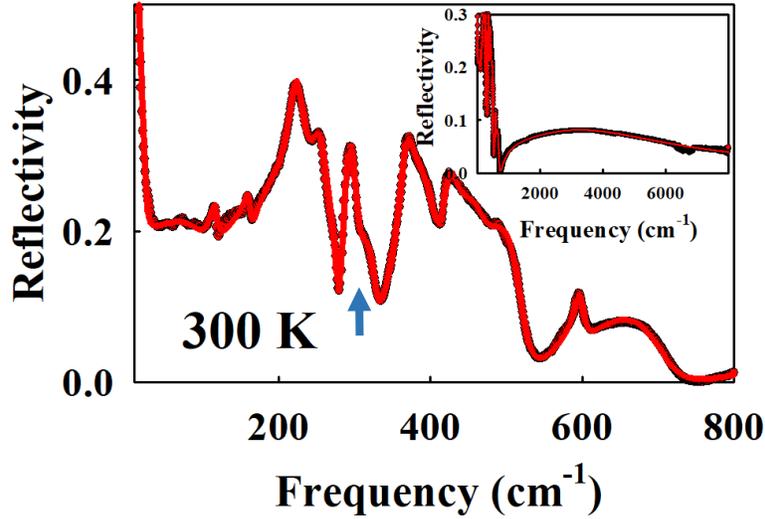

**Fig. 6** *h*-ErMnO$_3$ near normal reflectivity at 300 K, experimental: dots, full line: fit. Inset: full measured reflectivity range. Note that in the MIR part of these spectra to facilitate the fit we deleted second order phonons and weak Er$^{3+}$ crystal field substructure. Arrow points to the bands with anomalous temperature dependent profile. Fitting parameters are shown in Table SIV

## Table SIV

Dielectric simulation fitting parameters for *h*-ErMnO$_3$ reflectivity at 300 K. Bottom cells show the parameters used for the lower frequency Gaussian fit shown in Fig.S6. Phonon frequencies appearing in the polaron fits are indicated in bold italics with an asterisk.

| T (K) | $\varepsilon_\infty$ | $\omega_{TO}$ (cm$^{-1}$) | $\Gamma_{TO}$ (cm$^{-1}$) | $\omega_{LO}$ (cm$^{-1}$) | $\Gamma_{LO}$ (cm$^{-1}$) |
|---|---|---|---|---|---|
| 300 | 1.20 | 100.0 | 66.9 | 101.1 | 54.2 |
| | | 116.0 | 11.6 | 117.9 | 13.8 |
| | | 161.4 | 15.5 | 163.1 | 14.8 |
| | | 221.6 | 22.6 | 230.9 | 32.4 |
| | | 255.7 | 21.0 | 259.5 | 17.6 |
| | | 274.5 | 54.01 | 280.0 | 12.4 |
| | | ***288.2*** | ***12.0*** | ***307.3*** | ***36.5*** |
| | | ***\*324.9*** | ***22.6*** | ***328.9*** | ***18.9*** |
| | | 365.0 | 19.3 | 367.6 | 43.5 |
| | | 403.3 | 81.1 | 412.2 | 27.2 |
| | | 419.2 | 22.4 | 431.0 | 53.0 |
| | | 452.8 | 69.0 | 465.9 | 65.6 |
| | | 493.6 | 92.3 | 524.0 | 41.1 |
| | | 597.6 | 22.9 | 602.6 | 22.0 |
| | | *649.9 | 160.2 | 710.2 | 90.4 |
| | | 3852.4 | 6262.8 | 7446.3 | 6674.2 |
| | | A=542.6 | $\omega_{0G}$=6.47 | W=22.3 | |

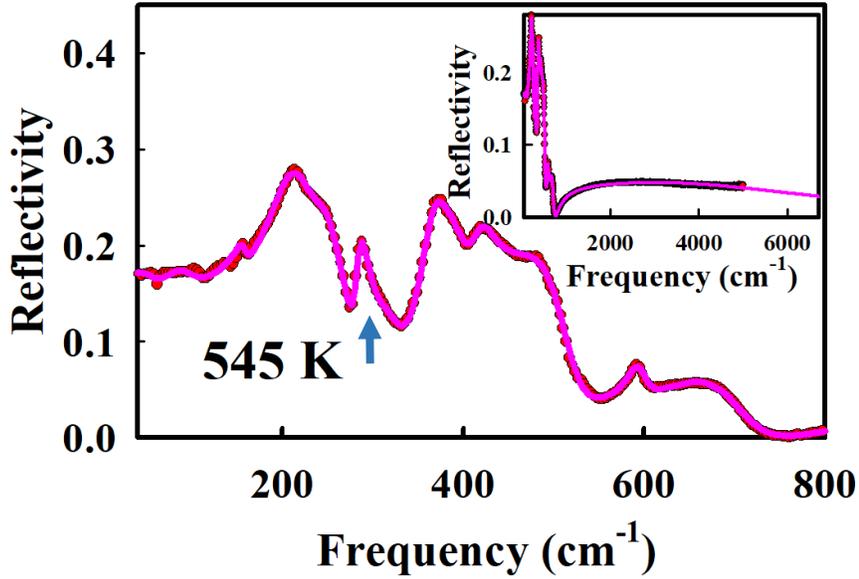

**Fig. S7** *h*-ErMnO$_3$ near normal reflectivity at 545 K, experimental: dots, full line: fit. Inset: full measured reflectivity range. Arrow points to the bands with anomalous temperature dependent profile. Fitting parameters are shown in Table SV.

## Table SV

Dielectric simulation fitting parameters for *h*-ErMnO$_3$ reflectivity at 545 K. Phonon frequencies appearing in the polaron fits are indicated in bold italics with an asterisk.

| T (K) | $\varepsilon_\infty$ | $\omega_{TO}$ (cm$^{-1}$) | $\Gamma_{TO}$ (cm$^{-1}$) | $\omega_{LO}$ (cm$^{-1}$) | $\Gamma_{LO}$ (cm$^{-1}$) |
|---|---|---|---|---|---|
| 545 | 1.07 | 58.9 | 35.3 | 60.2 | 34.2 |
| | | 94.4 | 46.5 | 97.7 | 50.0 |
| | | 157.6 | 15.3 | 158.6 | 15.6 |
| | | 209.9 | 44.3 | 224.4 | 73.2 |
| | | 248.7 | 40.0 | 251.5 | 56.3 |
| | | 277.7 | 157.5 | 280.9 | 22.1 |
| | | ***284.2*** | ***13.9*** | ***306.4*** | ***77.0*** |
| | | 315.0 | 39.8 | 319.0 | 51.2 |
| | | 363.6 | 32.8 | 366.2 | 71.2 |
| | | 405.7 | 102.5 | 410.1 | 42.0 |
| | | 414.8 | 33.6 | 424.1 | 62.9 |
| | | 428.2 | 97.3 | 450.4 | 159.8 |
| | | 497.2 | 100.2 | 516.4 | 55.2 |
| | | 593.9 | 29.0 | 597.7 | 28.1 |
| | | *681.8 | 168.7 | 711.8 | 83.7 |
| | | 4127.1 | 7139.7 | 6889.1 | 6745.8 |

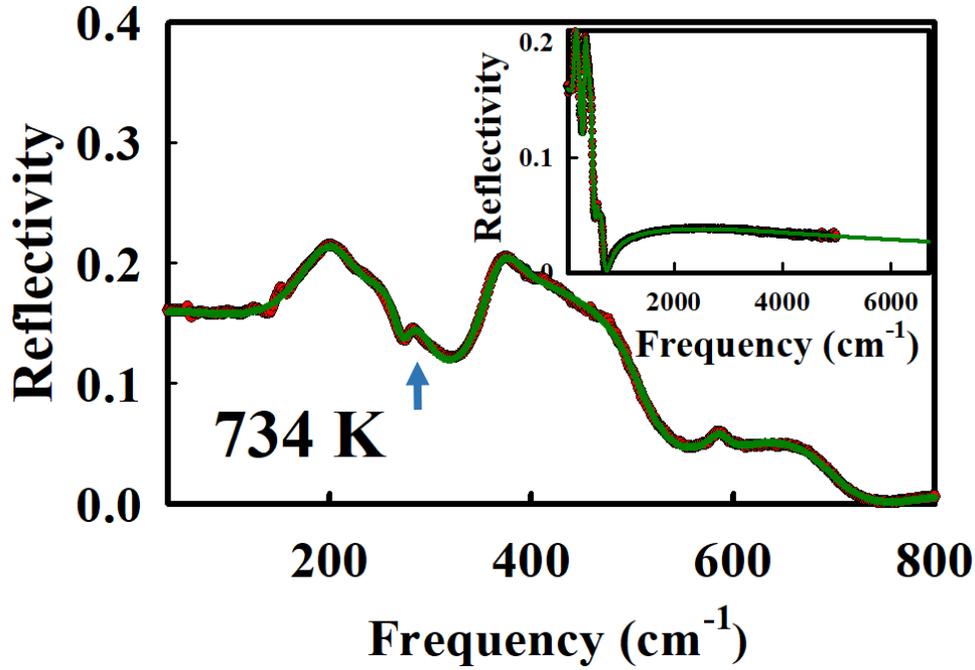

**Fig. S8** *h*-ErMnO$_3$ near normal reflectivity at 734 K, experimental: dots, full line: fit. Inset: full measured reflectivity range. Arrow points to the band with anomalous temperature dependent profile. Fitting parameters are shown in Table SVI

## Table SVI

Dielectric simulation fitting parameters for *h*-ErMnO$_3$ reflectivity at 734 K. Phonon frequencies appearing in the polaron fits are indicated in bold italics with an asterisk.

| T (K) | $\varepsilon_\infty$ | $\omega_{TO}$ (cm$^{-1}$) | $\Gamma_{TO}$ (cm$^{-1}$) | $\omega_{LO}$ (cm$^{-1}$) | $\Gamma_{LO}$ (cm$^{-1}$) |
|---|---|---|---|---|---|
| 734 K | 1.02 | 141.1 | 207.6 | 163.4 | 207.9 |
|  |  | 201.3 | 68.3 | 225.9 | 105.0 |
|  |  | 266.4 | 70.6 | 272-1 | 23.2 |
|  |  | ***\*275.0*** | ***29.0*** | ***300.3*** | ***102.6*** |
|  |  | 366.0 | 47.2 | 372.8 | 88.9 |
|  |  | 415.2 | 114.2 | 430.7 | 164.7 |
|  |  | 494.4 | 129.6 | 531.2 | 90.9 |
|  |  | 588.0 | 25.7 | 590.2 | 28.3 |
|  |  | \*674.5 | 178.1 | 739.4 | 54.6 |
|  |  | 3459.2 | 6335.7 | 1088.0 | 10131.1 |

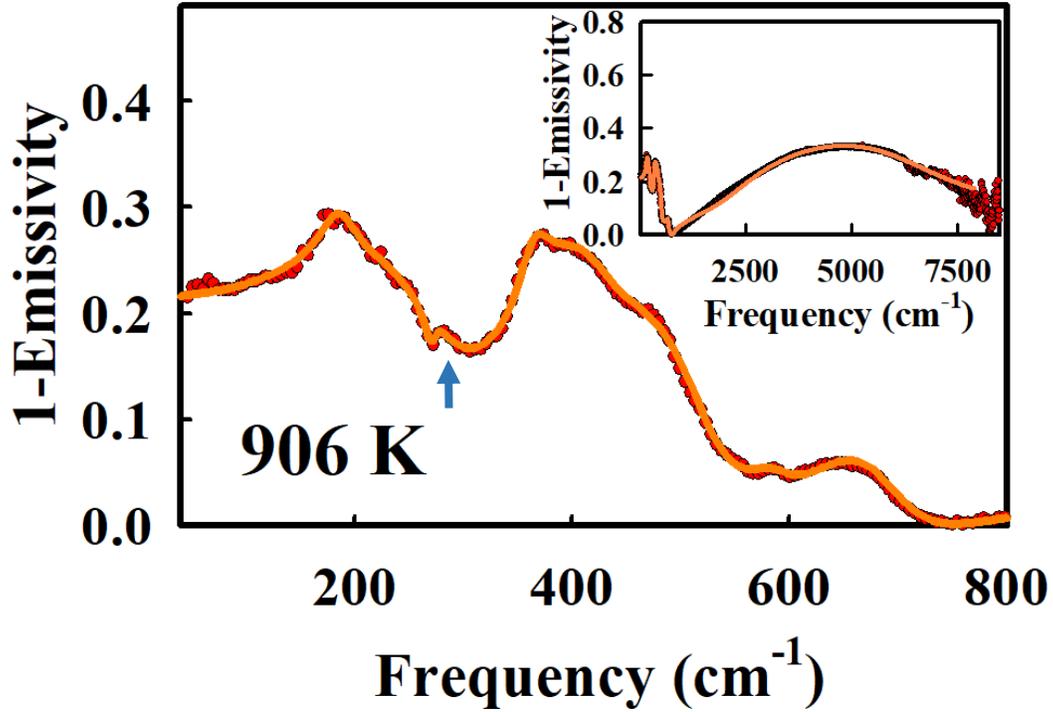

**Fig. S9** *h*-ErMnO$_3$ near normal 1-emissivity at 906 K, experimental: dots, full line: fit. Inset: full measured 1-emissivity range. Arrow points to the band with anomalous temperature dependent profile. Fitting parameters are shown in Table SVII

## Table SVII

Dielectric simulation fitting parameters for *h*-ErMnO$_3$ reflectivity at 906 K. Phonon frequency appearing in the bipolaron fit is indicated in bold italics with an asterisk.

| T (K) | $\varepsilon_\infty$ | $\omega_{TO}$ (cm$^{-1}$) | $\Gamma_{TO}$ (cm$^{-1}$) | $\omega_{LO}$ (cm$^{-1}$) | $\Gamma_{LO}$ (cm$^{-1}$) |
|---|---|---|---|---|---|
| 906 | 1.05 | 185.5 | 39.8 | 189.8 | 48.3 |
| | | 246.3 | 169.4 | 271.2 | 9.5 |
| | | ***271.8** | *10.0* | *288.9* | *118.6* |
| | | 363.1 | 35.2 | 367.6 | 45.3 |
| | | 412.5 | 101.5 | 427.7 | 79.8 |
| | | 443.2 | 167.5 | 529.2 | 97.1 |
| | | 587.8 | 51.1 | 591,9 | 47.5 |
| | | 654.5 | 143.0 | 703.9 | 87.6 |
| | | 2475.4 | 1891.5 | 6638.9 | 2902.5 |

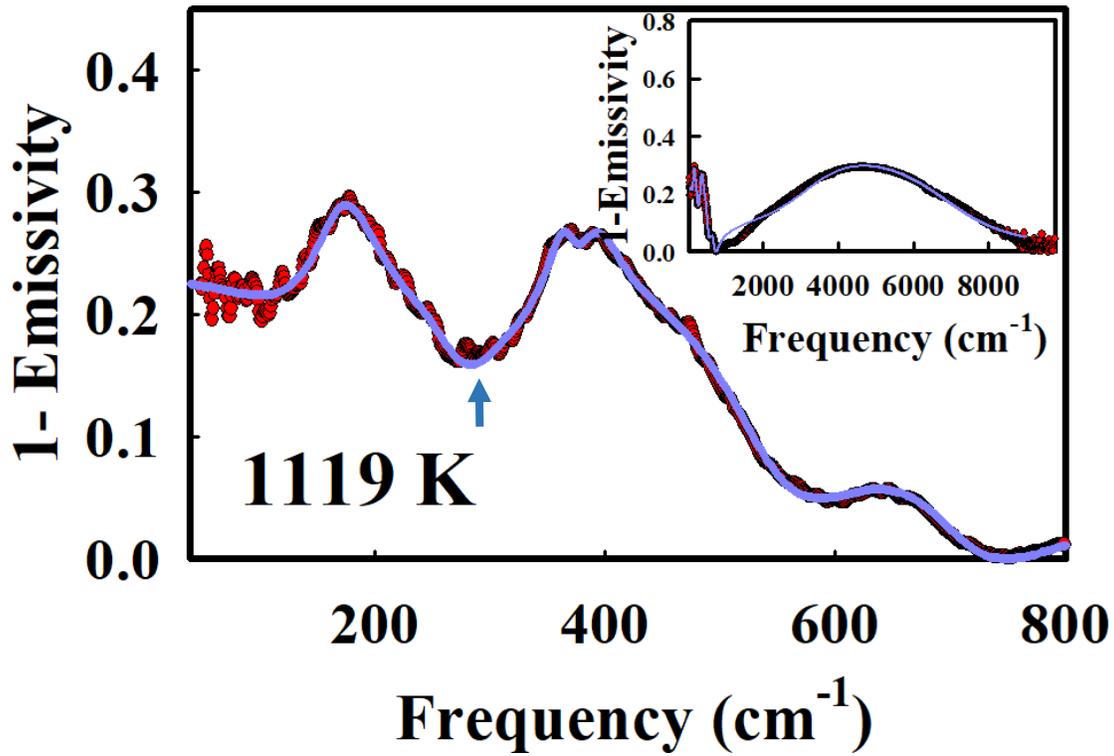

**Fig. S10** *h*-ErMnO$_3$ near normal 1-emissivity at 1119 K, experimental: dots, full line: fit. Inset: full measured 1-emissivity range. Arrow points to the band with anomalous temperature dependent profile. Fitting parameters are shown in Table SVIII

## Table SVIII

Dielectric simulation fitting parameters for *h*-ErMnO$_3$ reflectivity at 1119 K. Phonon frequency appearing in the bipolaron fit is indicated in bold italics with an asterisk.

| T (K) | $\varepsilon_\infty$ | $\omega_{TO}$ (cm$^{-1}$) | $\Gamma_{TO}$ (cm$^{-1}$) | $\omega_{LO}$ (cm$^{-1}$) | $\Gamma_{LO}$ (cm$^{-1}$) |
|---|---|---|---|---|---|
| 1119 | 1.12 | 164.7 | 252.1 | 169.7 | 187.3 |
|  |  | 172.4 | 53.7 | 182.9 | 104.1 |
|  |  | ***269.9** | *92.6* | *273.0* | *69.1* |
|  |  | 365.5 | 28.9 | 370.0 | 28.9 |
|  |  | 386.5 | 46.0 | 393.0 | 68.0 |
|  |  | 470.8 | 343.3 | 546.7 | 117.1 |
|  |  | 642.9 | 188.8 | 716.5 | 97.5 |
|  |  | 3450.1 | 2001.9 | 7220.1 | 2711.5 |

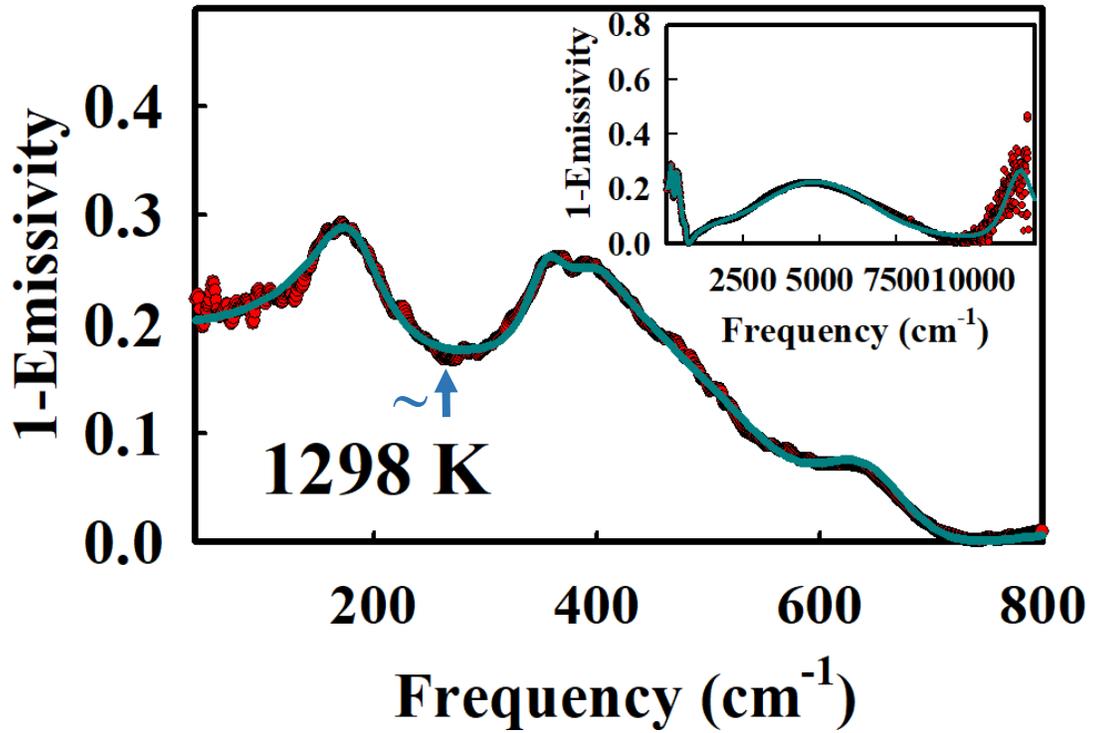

**Fig. S11** *h*-ErMnO$_3$ near normal 1-emissivity at 1298 K, experimental: dots, full line: fit. Inset: full measured 1-emissivity range. Arrow points to the band with anomalous temperature dependent profile. Fitting parameters are shown in Table SIX

## Table SIX

Dielectric simulation fitting parameters for *h*-ErMnO$_3$ reflectivity at 1298 K. Phonon frequency appearing in the bipolaron fit is indicated in bold italics with an asterisk.

.

| T (K) | $\varepsilon_\infty$ | $\omega_{TO}$ (cm$^{-1}$) | $\Gamma_{TO}$ (cm$^{-1}$) | $\omega_{LO}$ (cm$^{-1}$) | $\Gamma_{LO}$ (cm$^{-1}$) |
|---|---|---|---|---|---|
| 1298 | 1.13 | 176.1 | 64.8 | 200.2 | 101.0 |
| | | ***351.8*** | ***39.2*** | ***356.7*** | ***57.8*** |
| | | 391.8 | 61.7 | 394.9 | 84.0 |
| | | 488.8 | 510.9 | 578.3 | 165.9 |
| | | 625.7 | 110.0 | 679.6 | 105.8 |
| | | 3678.1 | 2509.9 | 7042.8 | 3092.7 |
| | | 10961.9 | 1250.0 | 11718.7 | 572.1 |

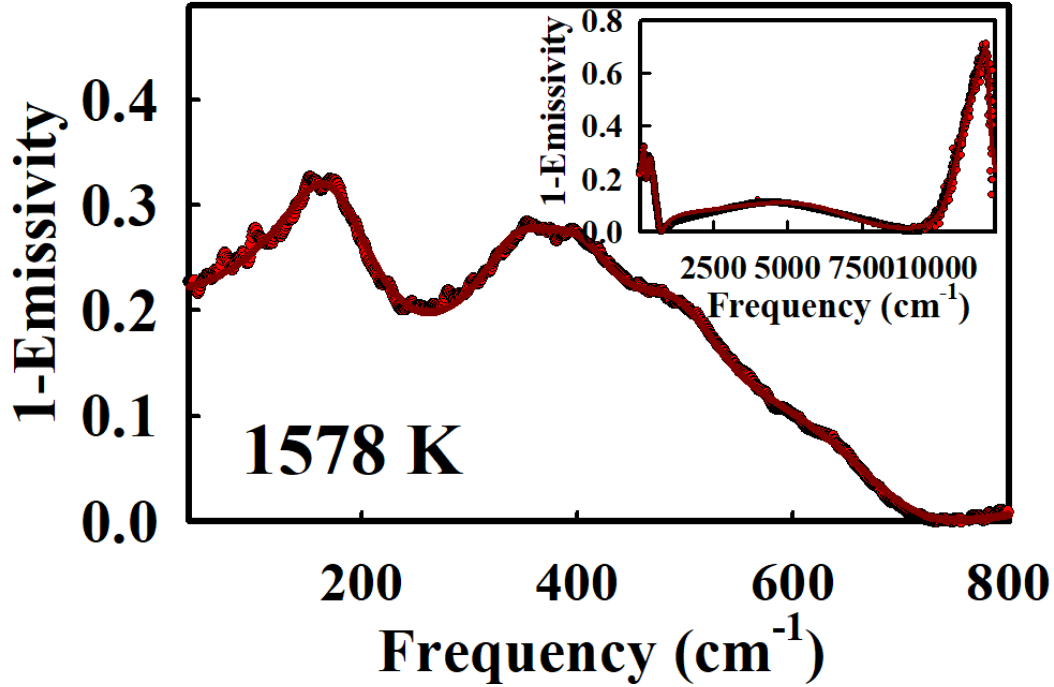

**Fig. S12** *h*-ErMnO$_3$ near normal 1-emissivity at 1578 K , experimental: dots, full line: fit. ; note that in this spectrum there is a slight increase intensity at the lowest frequencies signaling the onset of the insulating-metal transition. Inset: full measured 1-emissivity range. Fitting parameters are shown in Table SX

## Table SX

Dielectric simulation fitting parameters for *h*-ErMnO$_3$ reflectivity at 1578 K. The three phonon frequencies which average appears in the bipolaron fit are indicated with an asterisk

| T (K) | $\varepsilon_\infty$ | $\omega_{TO}$ (cm$^{-1}$) | $\Gamma_{TO}$ (cm$^{-1}$) | $\omega_{LO}$ (cm$^{-1}$) | $\Gamma_{LO}$ (cm$^{-1}$) |
|---|---|---|---|---|---|
| | | 165.1 | 80.0 | 213.2 | 142.9 |
| | | *339.6 | 85.7 | 362.5 | 124.4 |
| | | *397.7 | 82.1 | 411.0 | 101.5 |
| 1578 | 1.04 | *493.7 | 115,2 | 516.1 | 114.8 |
| | | 670.6 | 470.5 | 692.9 | 122.3 |
| | | 4733.1 | 5288.0 | 7440.2 | 3509.1 |
| | | 10492.2 | 924.5 | 11743.5 | 69.37 |

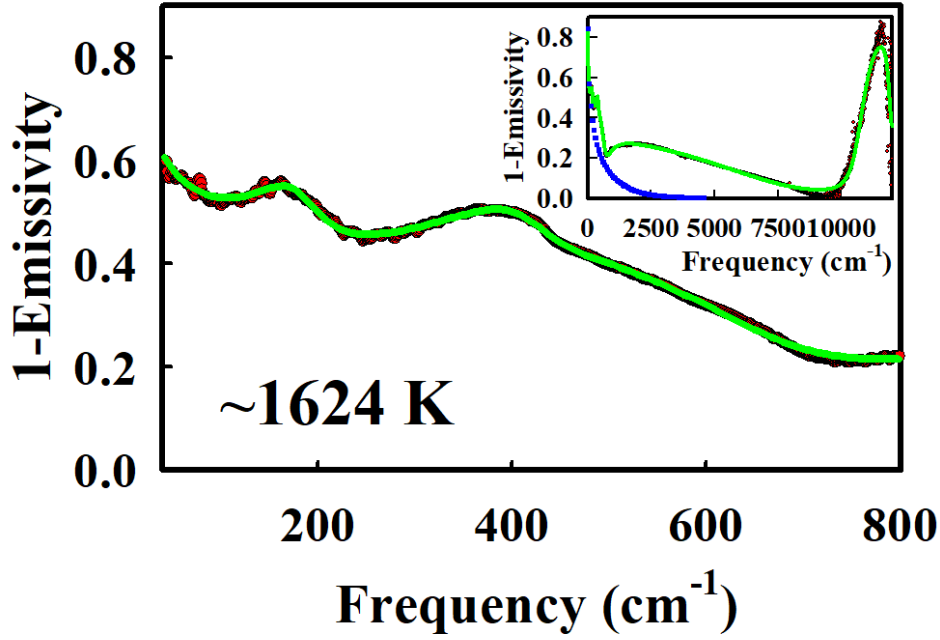

**Fig. S13** *h*-ErMnO$_3$ near normal 1-emissivity at ~1624 K showing the far infrared profile rise due to the Drude carriers, experimental: dots, full line: fit. Inset: full measured 1-emissivity range with the Drude term outlined at low frequencies. Fitting parameters are shown in Table SXI.

## Table SXI

Dielectric simulation fitting parameters for *h*-ErMnO$_3$ reflectivity at 1578 K.. Bottom cells contain the Drude parameters, eq(S5).

| (K) | $\varepsilon_\infty$ | $\omega_{TO}$ (cm$^{-1}$) | $\Gamma_{TO}$ (cm$^{-1}$) | $\omega_{LO}$ (cm$^{-1}$) | $\Gamma_{LO}$ (cm$^{-1}$) |
|---|---|---|---|---|---|
| | | 173.1 | 74.7 | 197.4 | 78.2 |
| | | 381.9 | 222.2 | 396.4 | 350.7 |
| | | 420.9 | 112.1 | 431,7 | 89.8 |
| | | 494.5 | 423.87 | 503.4 | 340.9 |
| ~1624 | 1.01 | 626.5 | 623.3 | 662.3 | 253.0 |
| | | 2276.9 | 5717.6 | 7648.2 | 6121.5 |
| | | 10657.8 | 726.4 | 11857.7 | 60.5 |
| | | | | $\omega_{pl}$ (cm$^{-1}$) | |
| | | | 2708.1 | 2103.2 | 3042.1 |